%% file: QShareTech.tex
\documentclass[sigconf]{acmart}

\usepackage{hyperref}
\usepackage{lipsum}
\usepackage{setspace}
\usepackage{booktabs, tabularx}
\usepackage{array, xspace, subfigure, url, xcolor, amsmath, bm, array,amsfonts, balance,multirow}

\definecolor{darkblue}{rgb}{0,0,0.5}
\hypersetup{
	pdftitle={Enabling Work-conserving Bandwidth Guarantees for Multi-tenant Datacenters via Dynamic Tenant-Queue Binding},
	pdfauthor={Anonymous}
	pdfkeywords={Work Conservation, Bandwidth Guarantees, Multi-tenant Datacenters},
	colorlinks=true,
	urlcolor=darkblue,
	linkcolor=darkblue,
	citecolor=darkblue
}

\usepackage[linesnumbered,ruled,vlined]{algorithm2e}
\usepackage[noend]{algpseudocode}

\SetCommentSty{mycommfont}

\newcommand{\ie}{\emph{i.e.,}\xspace}
\newcommand{\eg}{\emph{e.g.,}\xspace}
\newcommand{\sys}{QShare\xspace}

\newcommand{\first}{\textsf{(i)}\xspace}
\newcommand{\second}{\textsf{(ii)}\xspace}
\newcommand{\third}{\textsf{(iii)}\xspace}
\newcommand{\forth}{\textsf{(iv)}\xspace}

\newcommand{\es}{ElasticSwitch\xspace}

\newcommand{\paraspace}{\vspace{0.01in}}
\newcommand{\parab}[1]{\paraspace\noindent{\bf #1}}

\renewcommand\footnotetextcopyrightpermission[1]{} 
\usepackage[font={small,bf},skip=1pt]{caption}

\ccsdesc[500]{Networks~Cloud computing}
\ccsdesc[300]{Networks~Network resources allocation}

\begin{document}
\title{Enabling Work-conserving Bandwidth Guarantees for Multi-tenant Datacenters via Dynamic Tenant-Queue Binding}
\titlenote{The initial work is published in IEEE INFOCOM 2018~\cite{infocomversion}.}
\author{Zhuotao Liu$^{1,2}$ Kai Chen$^{1}$ Haitao Wu$^{3}$ Shuihai Hu$^{1}$ Yih-Chun Hu$^{2}$ Yi Wang$^4$ Gong Zhang$^5$}
\titlenote{This work is done in part when Zhuotao Liu was interning at SING Lab, HKUST.}
\affiliation{
	\institution{$^{1}$Hong Kong University of Science and Technology}
}
\affiliation{
$^{2}$University of Illinois at Urbana-Champaign
}
\affiliation{
	\institution{$^{3}$Google $^{4}$Southern University of Science and Technology $^5$Huawei Future Network Theory Lab}
}
\affiliation{\{zliu48,yichun\}@illinois.edu, \{kaichen,shuaa\}@cse.ust.hk}

\input{abs}
\maketitle

\input{intro}

\input{motivation}

\input{sys_overview}
\input{tenant_placement}
\input{queue_binding}

\input{implementation}

\input{evaluation}
\input{related}

\input{conclusion}

\bibliographystyle{acm}
\bibliography{paper} 

\newpage
\balance
\input{appendix}

\end{document}

%% file: abs.tex
\begin{abstract}\label{sec:abs}
Today's cloud networks are shared among many tenants. Bandwidth guarantees and work conservation 
are two key properties to ensure predictable performance for tenant applications and high network utilization for providers. 
Despite significant efforts, very little prior work can \emph{really} achieve both properties simultaneously even some of them claimed so. 

In this paper, we present \sys, an in-network based solution 
to achieve bandwidth guarantees and work 
conservation simultaneously. \sys leverages weighted fair queuing on commodity 
switches to slice network bandwidth for tenants, and solves the 
challenge of queue scarcity through balanced 
tenant placement and dynamic tenant-queue binding. \sys is readily 
implementable with existing switching chips. 
We have implemented a \sys prototype and evaluated it via both testbed 
experiments and simulations. Our results show that \sys ensures bandwidth 
guarantees while driving network utilization to over $91\%$ even under 
unpredictable traffic demands.
\end{abstract}

%% file: intro.tex
\section{Introduction}\label{sec:intro}
Sharing the network of multi-tenant datacenters has been a critical theme for public clouds. The two primary objectives, among others, are bandwidth guarantees and work conservation. Bandwidth guarantees ensure predictable lower bound network performance for tenant applications. Recent studies show that, without bandwidth guarantees, network performance can experience 5x or more variations, leading to poor application performance~\cite{oktopus}. Work conservation enables a tenant to use spare bandwidth beyond its minimum guarantee to further improve its application performance as well as boost provider network utilization. Given that datacenter traffic is bursty in nature and that the average network utilization is low~\cite{facebook, dc_mesaure,imc2010}, work conservation can deliver over 10x additional bandwidth to a tenant VM upon its minimum guarantee~\cite{elasticswitch}.

However, it is hard to achieve both bandwidth guarantees and work conservation simultaneously. Prior works such as Oktopus~\cite{oktopus} and SecondNet~\cite{secondnet} can achieve bandwidth guarantees, but they are not work-conserving. Seawall~\cite{seawall} and NetShare~\cite{netshare} achieve work conservation, but they do not provide bandwidth guarantees (more details in $\S$\ref{sec:related}).

ElasticSwitch~\cite{elasticswitch} takes the first step toward achieving both properties at the same time. It is an endhost based solution that first needs to translate per-VM hose-model bandwidth guarantees into VM-to-VM pair rate limiters (referred as Guarantee Partitioning, GP), and then dynamically allocates spare bandwidth to these VM pairs to achieve high utilization (referred as Rate Allocation, RA). However, this approach confronts two challenges: \first as tenant applications are typically agnostic to network operators, it is difficult for GP to estimate each tenant's traffic matrix (including VM-to-VM communication patterns and per VM-pair demand), affecting both bandwidth guarantees and work conservation;  \second to detect spare bandwidth, RA needs to probe the network by increasing rates, which causes a tradeoff between accurately providing bandwidth guarantees and being work conserving~\cite{elasticswitch,trinity}: a conservative RA sacrifices work conservation, while an aggressive RA affects other tenants' bandwidth guarantees (see our experiments in $\S$\ref{sec:appendix}).

Trinity~\cite{trinity} moves one step further to complement \es with simple in-network support. It uses two priority queues in switches to segregate and prioritize the bandwidth guarantee traffic over work conservation traffic, so that aggressive RA of a tenant does not affect bandwidth guarantees of others. While Trinity solves the second challenge of \es,  it still suffers from the more fundamental challenge of executing GP without prior knowledge of tenant traffic matrix.  Further, it introduces other issues such as packet reordering and starvation due to traffic segregation and priority queuing. 

As a result, prior solutions, essentially, do not achieve both goals in a sufficient manner. To give some sense, our testbed experiments show that without prior knowledge of tenant traffic matrix, state-of-the-art solutions relying on GP fail to achieve good work-conversation (given bandwidth guarantees are satisfied), which, for instance, causes 2x long flow completion times (FCTs) for tenant applications compared to our proposed solution.

Motivated by this, in this paper, we propose \sys, a comprehensive in-network solution to address the above challenges so as to achieve both goals in a sufficient manner. Instead of using two priority queues to segregate traffic for two different types, \sys directly leverages multiple weighted fair queues (WFQs) to slice network bandwidth for tenants. This ensures that \first bandwidth guarantees are achieved through proper queue weight configuration and tenant placement rather than endhost rate limiters, thus relieving us of GP; \second the network link is driven to full utilization instantly as long as one tenant has sufficient demand; \third no matter how aggressively a tenant transmits, bandwidth guarantees of other tenants are not affected as they are served in separate weighted queues; \forth no packet reordering or starvation arises. While promising, \sys faces a practical challenge of queue scarcity---the number of queues on a commodity switch port (typically 8) can be less than the number of tenants served by this port (see \S\ref{sec:evaluation:queue_scarcity} for detailed analysis in large scale datacenters).

To address this challenge, we make the following observation: although the total number of embedded tenants associating with a port may be large, during a short time interval (\eg a few seconds), the number of concurrent tenants whose traffic demands exceed their bandwidth guarantees is small. This is also reflected by the measurement results in production datacenters, where the average link utilization is low~\cite{facebook, dc_mesaure,imc2010}. Thus, to support more tenants with limited queues, \sys dynamically assigns dedicated queues for tenants with higher demands than their guarantees, while serving the tenants whose current demands are relatively low than their bandwidth guarantees in a shared queue altogether. 

\sys mainly contains two modules: a balanced tenant placement module and a dynamic tenant-queue binding module ($\S$\ref{sec:system_overview}). The tenant placement module is responsible for allocating network resources to tenants to provide bandwidth guarantees. To facilitate the dynamic queue allocation for embedded tenants, our placement module also aims to balance the usage of switch ports among tenants to avoid overwhelming certain ports. The tenant-queue binding module then takes into account the traffic demands of tenants and their payment factors to dynamically distribute queue resources among tenants.

We implement a prototype of \sys with ${\sim}2000$ lines of code (\textsf{C} for Linux kernel space and \textsf{Python} for user space), and perform extensive evaluations on testbed and via simulations. Our evaluation results suggest that:
\begin{itemize}
\item Without sacrificing bandwidth guarantees, \sys achieves \first \emph{perfect} work conservation given correct prediction on demand trends (not the exact traffic matrix), and \second over $91\%$ link utilization given completely  \emph{unpredictable} demands. 

\item Given the above desirable properties, \sys  significantly benefits applications, for instance, by reducing their flow completion times (FCTs) by up to $50\%$ compared with the state-of-the-art~\cite{elasticswitch,trinity}. 

\item With production datacenter settings, \sys can assign dedicated queues to ${\sim}90\%$ of all embedded tenants even when the datacenter is fully reserved, yielding at least $3\times$ throughput gain over their bandwidth guarantees and better efficiency in link utilization. 
\end{itemize}

%% file: motivation.tex
\section{Background and Motivation}\label{sec:motivation}

\subsection{Background}\label{subsec:background}
In multi-tenant datacenters~\cite{oktopus,secondnet,tag,netlord}, a conceptually centralized \emph{tenant manager} with global view of the datacenter state is responsible for managing all tenants, including tenant embedding, routing updates, logging,failure handling \& recovery and so forth. By designing various components for the tenant manager, datacenter operators are able to achieve self-interested goals, such as accommodating more tenants and achieving efficient resource utilization. \sys can be viewed as a newly designed component in the tenant manager to simultaneously accomplish the following two desirable properties: bandwidth guarantees and work conservation.

\begin{figure}[t]
	\centering
	\mbox{
		\subfigure[Hose model for B.G.\label{fig:hose:a}]{\includegraphics[scale=0.47]{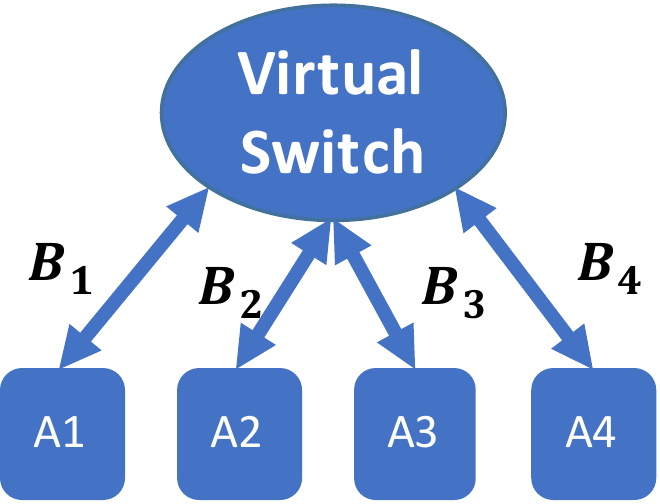}}~
		\subfigure[Bandwidth reservation\label{fig:hose:b}]{\includegraphics[scale=0.47]{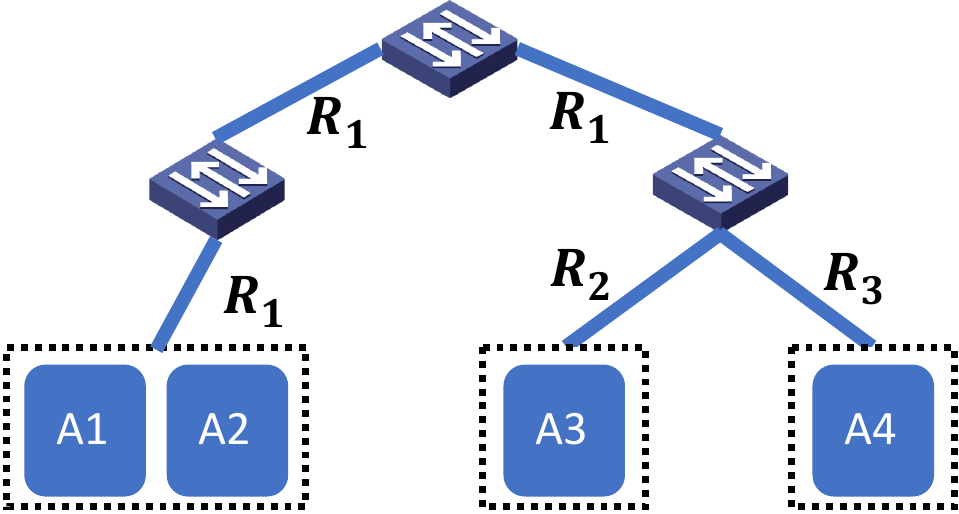}}
	}
	\caption{Figure~\ref{fig:hose:a} shows the tenant (VM) bandwidth guarantees defined in a symmetric hose model. Figure~\ref{fig:hose:b} shows the 
		reserved bandwidth on each link after embedding the tenant: $R_1{=}\min\{B_1{+}B_2, B_3{+}B_4\}$; $R_2 {=} \min\{ B_3, B_1 {+} B_2 {+} B_4\}$; 
		$R_3 {=} \min\{ B_4, B_1 {+} B_2 {+} B_3\}$.
	}
	\label{fig:hose}
\end{figure}

Network bandwidth guarantees are preferable properties in cloud computing to offer tenants predicable performance. A typical way to model bandwidth guarantees is using the hose model~\cite{hose,oktopus,faircloud,eyeq,gatekeeper,elasticswitch,tag}. As an illustrative example, Figure~\ref{fig:hose:a} shows a tenant \texttt{A}'s bandwidth guarantees defined in a hose model, and Figure~\ref{fig:hose:b} illustrates the reserved bandwidth on each physical link to satisfy the bandwidth guarantees after tenant embedding. For simplicity, a symmetric hose model is plotted in Figure~\ref{fig:hose}. Providing accurate bandwidth guarantees for VMs that can use multiple paths is an open problem since it requires a \emph{perfect} load balancer to accurately distribute each VM's traffic over multiple paths such that the sum of guarantee on each path equals to the  total amount of guaranteed bandwidth. As a result, prior proposals for providing bandwidth guarantees are either within the scope of tree-based network topology~\cite{secondnet, oktopus, elasticswitch} or confining each tenant's traffic within a tree in multi-path network topologies~\cite{tag,silo,OpReduce}. \sys belongs to the second category as typical datacenters (\eg Clos~\cite{clos,fattree}) are built with path redundancy. \sys, however, can still fully utilizes the redundant network links via balanced tenant placement.

Work conservation is desired for achieving efficient resource utilization. Formally, in the context of multi-tenant datacenters, work conservation is defined as follows: for any link $L$ in the network, as long as there exists at least one tenant that has packets to send along link $L$, $L$ cannot have spare bandwidth~\cite{faircloud}. We note that work-conservation does not guarantee that there are no idle links in the network. Idle links may exist due to the lack of traffic demands or high-demanded tenants are bottlenecked by other links.

\subsection{State-of-the-Art Solutions}\label{sec:state_of_the_art}
ElasticSwitch~\cite{elasticswitch} makes the first attempt to achieve work-conserving bandwidth guarantees. It is an end-host based solution composed of two modules: a Guarantee Partitioning (GP) module that divides VM $X$'s hose-model guarantee into guarantees to/from each other VM that $X$ communicates with, and a Rate Allocation (RA) module that assigns spare bandwidth to these VM pairs to achieve high network utilization. However, it suffers from the following two key challenges.

First, since the traffic matrix (TM) among the VMs of a tenant is typically agnostic to cloud providers, GP has to gradually  learn each VM-pair's demand via periodic source-destination VM coordination and throughput measurement. Whenever the TM changes, GP needs to re-estimate the TM even if per-VM demand remains the same (see illustrative example in \S \ref{sec:appendix}). Given highly bursty and dynamic TM in datacenters, it is challenging for the GP to capture the real communication pattern and estimate the TM correctly, especially considering that tens of thousands of VMs can produce billions of VM pairs. Performing GP in a dynamic context without the prior knowledge of tenant TM imposes a fundamental challenge for \es since it affects both bandwidth guarantees and work conservation. Further, at such scale, the overhead of maintaining these VM-pair rate limiters at hypervisors is non-negligible~\cite{elasticswitch}.

Second, RA in \es~\cite{elasticswitch} aims to grab available network bandwidth beyond the provided guarantees. It probes the network by increasing rates, detects congestion via packets losses or ECN, and then allocates the spare bandwidth to VM pairs in max-min fashion following weighted TCP algorithms~\cite{seawall, seawall-like}. As mentioned in \cite{elasticswitch,trinity}, it has a tradeoff between accurately providing bandwidth guarantees and being work-conserving: aggressive RA could affect other tenants' guarantees whereas conservative RA ends up with bandwidth waste. In practice, RA's performance  depends on the parameter choice and system tuning.

Trinity~\cite{trinity} moves one step further to complement the endhost based ElasticSwitch with simple in-network support. It exploits two priority queues in switches to segregate and prioritize the bandwidth guarantee traffic over work conservation traffic. As a result, VMs can send work-conservation traffic more aggressively than \es with less affect on bandwidth guarantees of other tenants. Thus, Trinity achieves work conservation in a static context, \ie the demand of each VM-pair (\ie the TM) is a priori knowledge. However, it still suffers from the fundamental challenge of executing GP in dynamic context since it still needs to translate per-VM house-model bandwidth guarantees into VM-pair rate limiters on hypervisors. Further, since network traffic is segregated and served with strict priorities, Trinity raises packet reordering and starvation issues in practice. 

We do preform detailed experiments and analysis to quantify these limitations. Please see detailed results in \S\ref{sec:appendix}. Motivated by this, we propose \sys to address these challenges.

%% file: sys_overview.tex
\section{\sys Overview}\label{sec:system_overview}

\sys is a comprehensive in-network solution to address the above challenges. Instead of using two priority queues to segregate traffic for two different types, \sys leverages multiple weighted fair queues (WFQ) (WFQ is emulated by WRR on some switches) to slice network bandwidth for tenants. This enables \sys to provide tenant-level bandwidth guarantees and work conservation (instead of rigid VM-to-VM pair level as in both \es~\cite{elasticswitch} and Trinity~\cite{trinity}), thus leaving tenant applications full flexibility to use their allocated bandwidth as needed. We note that such tenant-level bandwidth guarantees are also used in~\cite{end-to-end, oktopus}, but they fail to achieve work conservation.

\begin{figure}[t]
	\centering
	\mbox{
		\subfigure[In-network Support\label{fig:no_tradeoff:a}]{\includegraphics[scale=0.25]{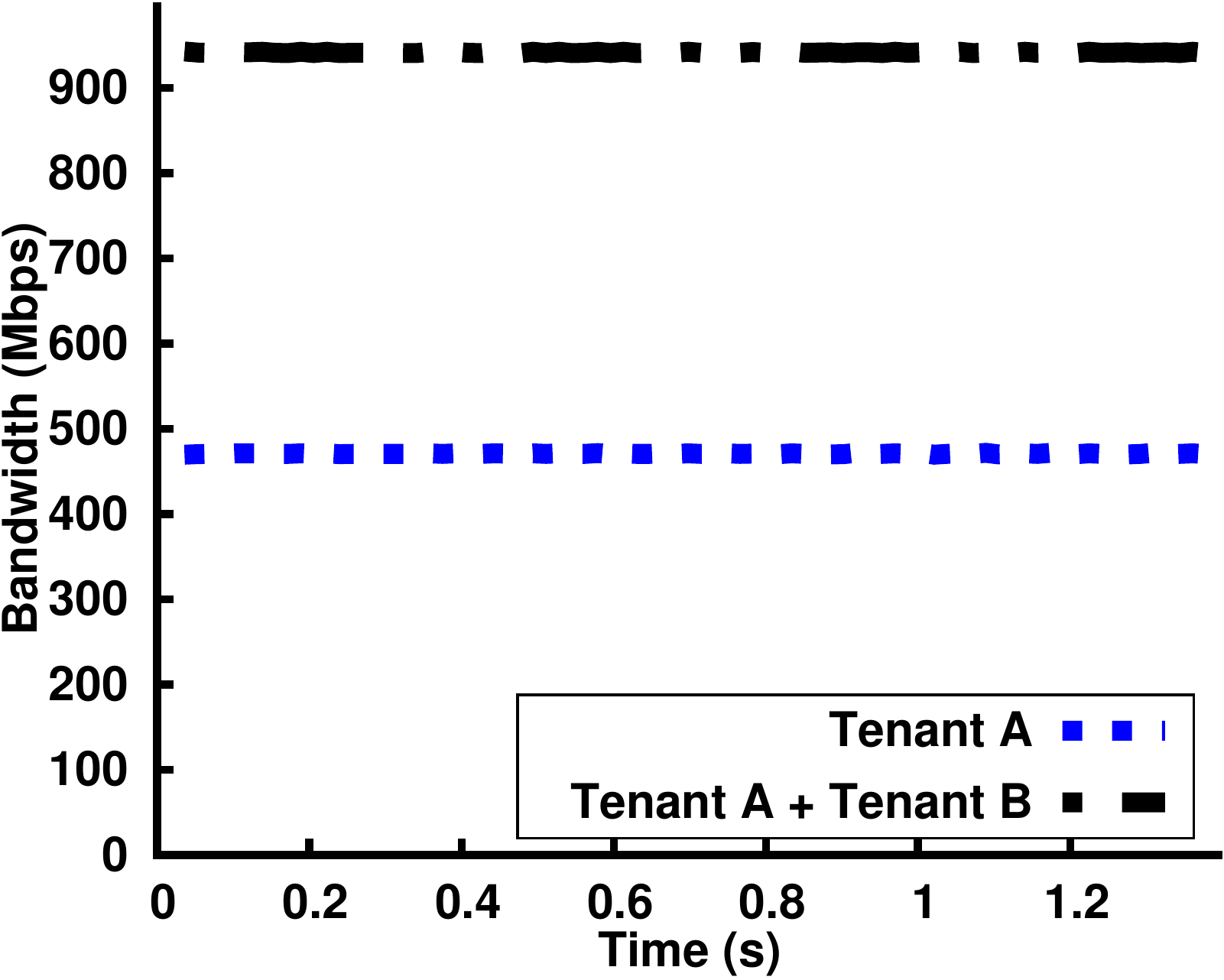}}~~
		\subfigure[Pure endhost-based\label{fig:temp_tradeoff:b}]{\includegraphics[scale=0.25]{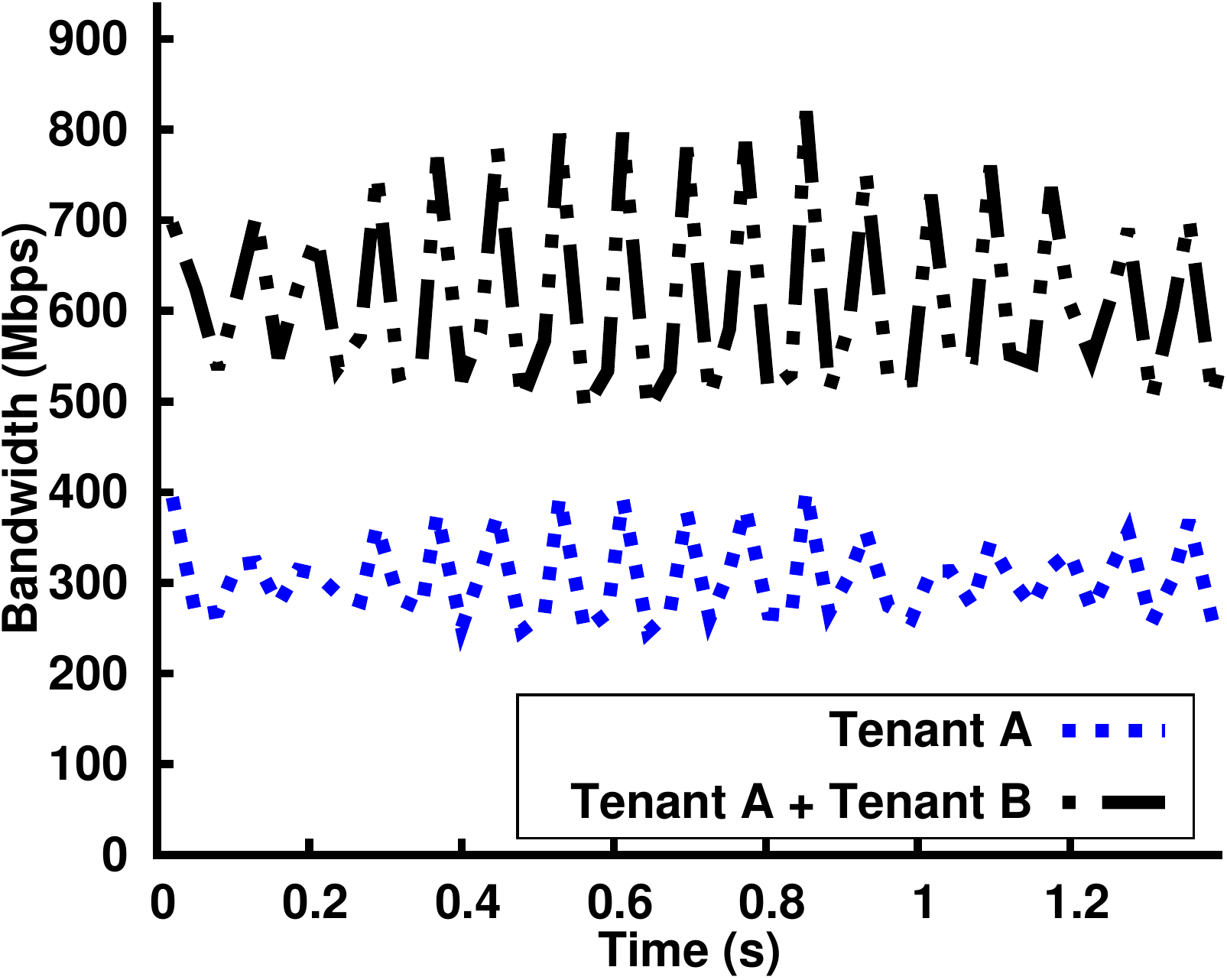}}
	}
	\caption{Compared with pure endhost-based solution, \sys achieves perfect (no tradeoff) work-conserving bandwidth guarantees via in-network WFQ support.}
	\label{fig:no_tradeoff}
\end{figure}

\begin{figure*}[t]
  \centering
  \mbox{
	\subfigure[In-network support\label{fig:system_illu:a}]{\includegraphics[scale=0.415]{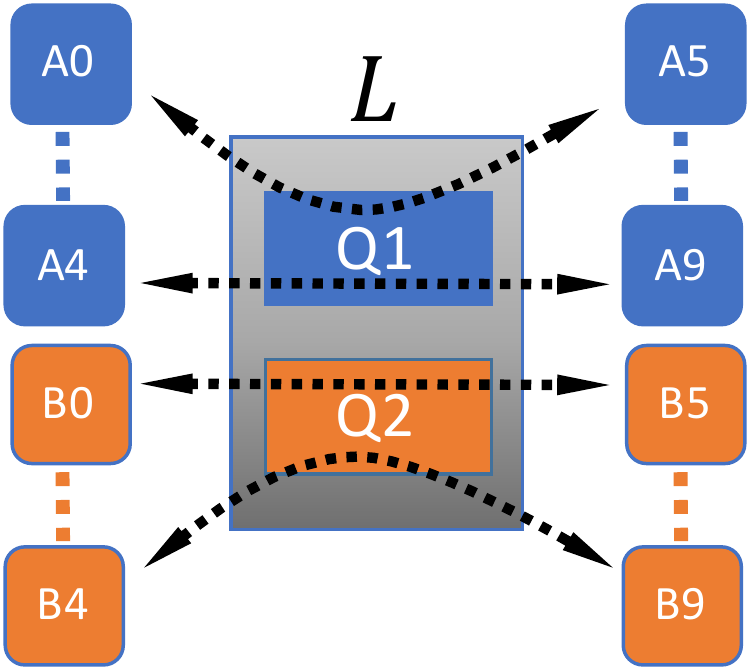}}\quad
	\subfigure[Balanced tenant placement\label{fig:system_illu:b}]{\includegraphics[scale=0.32]{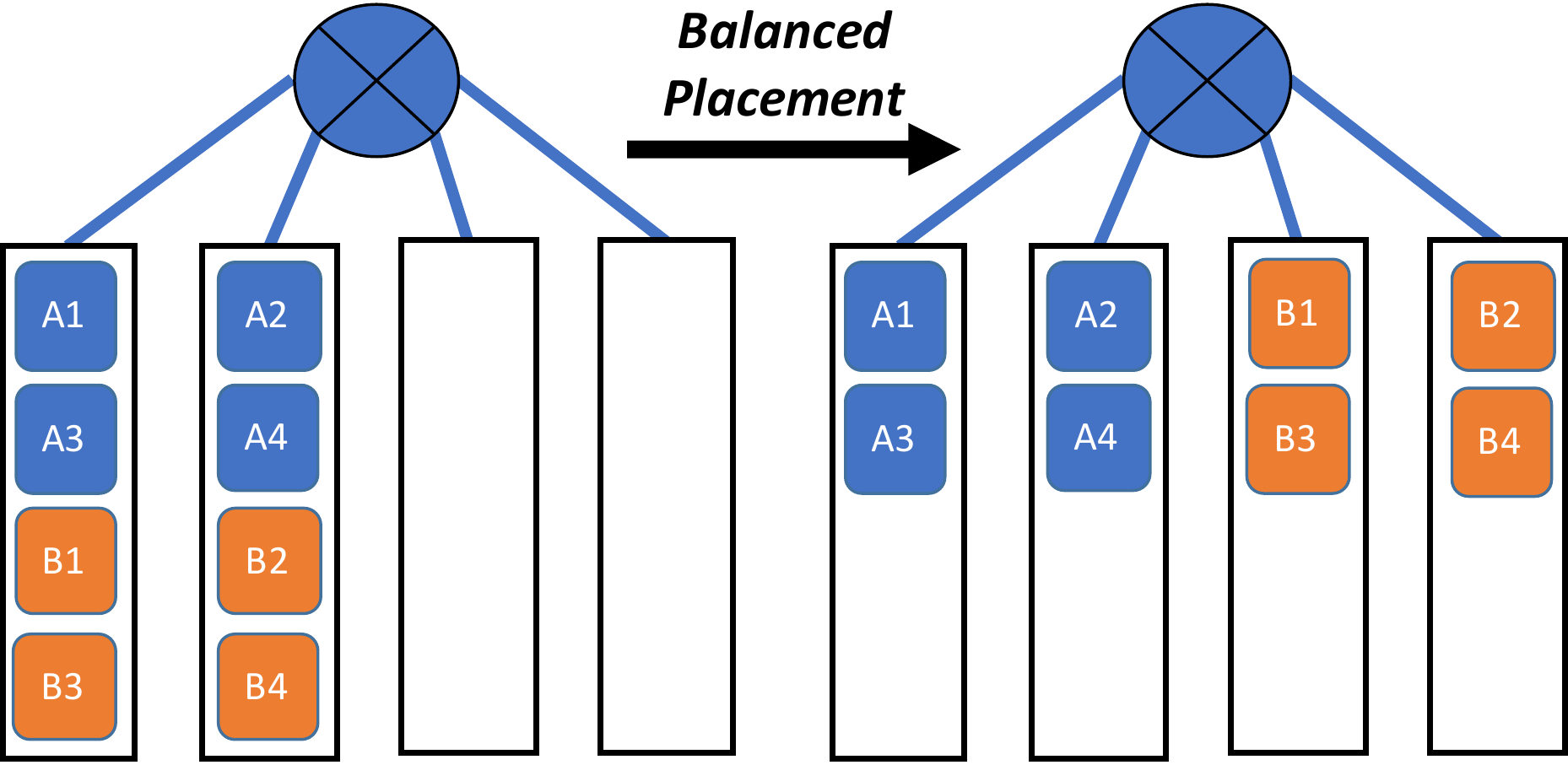}}\quad
	\subfigure[Tenant-queue allocation\label{fig:system_illu:c}]{\includegraphics[scale=0.43]{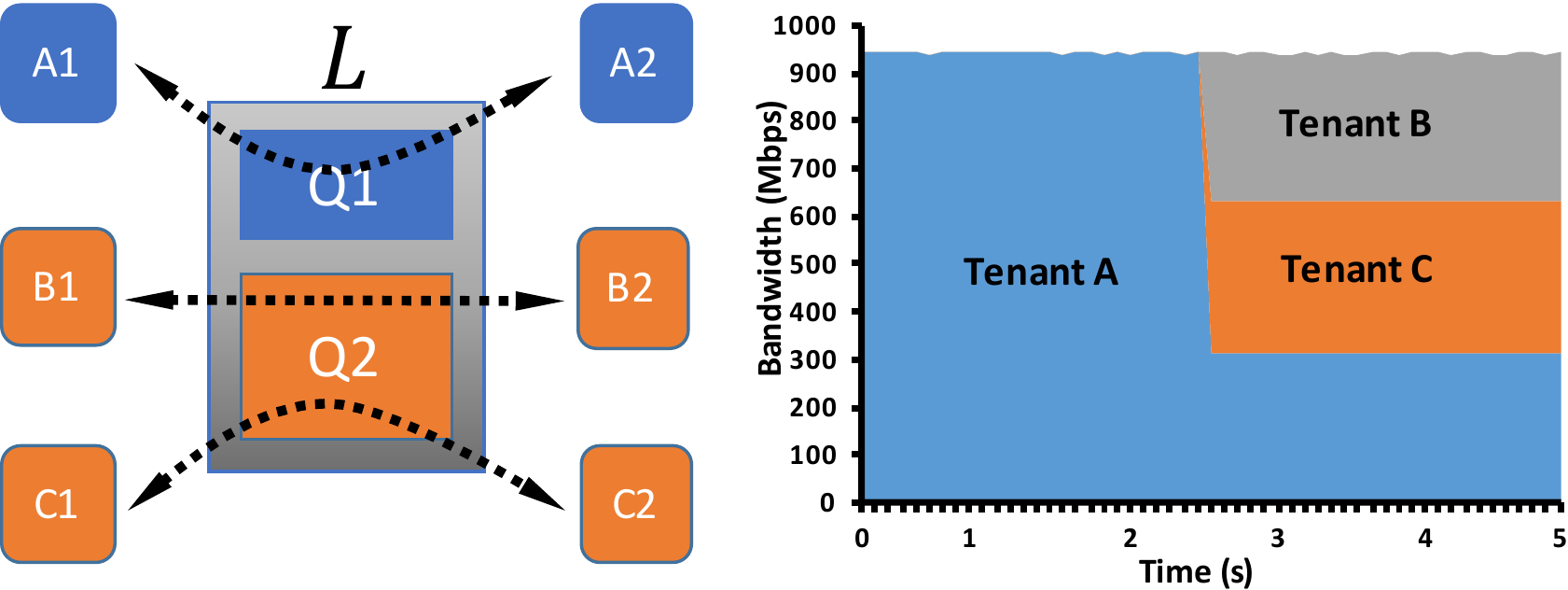}}
  }
  \caption{Three illustrative examples for \sys's design. Figure~\ref{fig:system_illu:a} shows that \sys incorporates the in-network WFQ support. Thus, tenants \textsf{A} and \textsf{B} sharing the link $L$ are served in $2$ separate weighted queues. Figure~\ref{fig:system_illu:b} shows that \sys's tenant placement algorithm balances the usage of switch ports among the embedded tenants to avoid overwhelming certain ports. Figure~\ref{fig:system_illu:c} plots an illustrative example to show \sys's key design: assigning a dedicated queue for the high-demanded tenant \textsf{A} achieves perfect work conservation without estimating \textsf{A}'s TM, and meanwhile, tenants \textsf{B} and \textsf{C} served in the shared queue immediately receive their guaranteed bandwidth once they become active.}
\label{fig:system_illu}
\end{figure*}

\subsection{In-Network WFQ Support}\label{sec:in_network_support}
WFQ on commodity switches offers desirable in-network support for achieving work-conserving bandwidth guarantees. We use the following toy experiment to demonstrate its benefit. We place two tenants \textsf{A} and \textsf{B}, both provisioned with $10$ VMs, on our testbed. Each tenant's VMs are evenly distributed across two racks connected by a core link with $1$Gbps capacity. As \textsf{A} and \textsf{B} share the core link $L$, their flows are served in two separate WFQ queues whose weights are configured proportionally to their guaranteed bandwidth on the link (Figure \ref{fig:system_illu:a}). 

Consider a case where both \textsf{A} and \textsf{B} adopt the same symmetric hose model, in which each VM is guaranteed $50$Mbps bandwidth in its hose model. Thus, both tenants have $250$Mbps guaranteed bandwidth  on the core link. To generate traffic, each VM in one rack is configured to communicate with randomly selected VMs, using our client/server program described in \S \ref{sec:evaluation}. Each VM's demand and its communication pattern are completely random. Only intra-tenant communication is considered. We measure the amount of core-link bandwidth utilized by each tenant. As plotted in Figure~\ref{fig:no_tradeoff:a}, without relying on any TM estimation, WFQ enables \sys to achieve perfect work-conserving bandwidth guarantees without imposing packet reordering or starvation issues. We repeat the experiment using self-implemented prototype of \es. As illustrated in Figure~\ref{fig:temp_tradeoff:b}, we notice a significant gap (over $300$Mbps) between the aggregate bandwidth of \textsf{A} \& \textsf{B} and the link capacity, \ie over 60\% of the unreserved bandwidth is wasted. We are aware that \es's performance depends on parameter settings. We consider different settings in \S \ref{sec:appendix}.

\subsection{Design Overview}
The key challenge of \sys is to address the problem of queue scarcity: the number of queues on each switch port (typically $8$) can be less than the number of tenants served by this port, and therefore we cannot allocate a dedicated queue for each tenant. Thus, to achieve the benefits for in-network WFQ support, \sys designs two modules: a balanced tenant placement module and a dynamic tenant-queue binding module. 

The placement module first seeks to provision tenant network to ensure bandwidth guarantees for each tenant. Further, it \emph{balances} the usage of switch ports among tenants to reduce the stress of performing the dynamic queue allocation in the binding module. For instance, if both placements in Figure~\ref{fig:system_illu:b} satisfy bandwidth guarantees, \sys prefers the one on the right side since the switch ports (and their queues) are more evenly utilized by the tenants. 

The tenant-queue binding module dynamically assigns dedicated queues to tenants whose demands are higher than their guaranteed bandwidth, and meanwhile serves all the low-demanded tenants in a shared queue (they may employ \es-like rate allocation to improve the worst case performance, as explained below). As a result, tenants in dedicated queues can burst their traffic in arbitrary communication patterns  without affecting other tenants. This design is the key to avoid the challenging GP and to eliminate the tradeoff between bandwidth guarantees and work conservation in the endhost based solutions~\cite{elasticswitch,trinity}. We perform an experiment to demonstrate this. Consider that three tenants \textsf{A}, \textsf{B} and \textsf{C} compete on a link $L$ with $1$Gbps capacity, shown in Figure~\ref{fig:system_illu:c}. Each tenant has $300$Mbps guarantee on $L$. Assume for now $L$ only has  $2$ WFQ queues. Suppose tenant \textsf{A} is high-demanded so that \sys assigns it a dedicated queue (with weight $1$) whereas \textsf{B} and \textsf{C} share a common queue (with weight $2$). When only \textsf{A} is active, it can fully utilize the link capacity with arbitrary communication pattens, achieving work conservation. Further, tenants \textsf{B} and \textsf{C} are not overwhelmed by \textsf{A}, and immediately receive their guaranteed bandwidth once becoming active.

\begin{figure}[t]
	\centering
	\mbox{
		\subfigure{\includegraphics[scale=0.4]{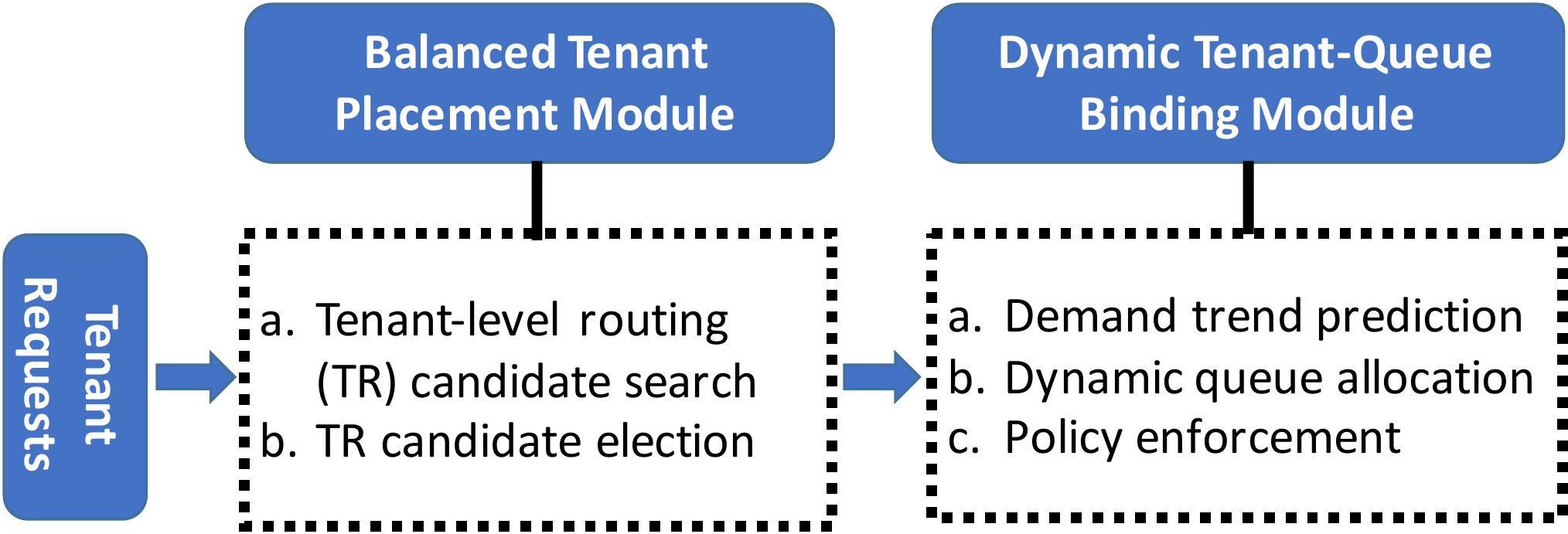}}
	}
	\caption{The architecture of \sys.}
	\label{fig:sys_arch}
\end{figure}

The key challenge of our dynamic binding mechanism is how to assign right tenants dedicated queues since traffic demand is dynamic. \sys addresses the challenge as follows. First, rather than predicting traffic matrix for each tenant as proposed in~\cite{elasticswitch,trinity}, \sys's demand prediction relies on only a scalar metric (detailed in \S \ref{sec:queue_binding:demand}) of each tenant, which greatly reduces stress of prediction. Second, to improve the worst case performance when traffic demand prediction is inaccurate and high-demanded tenants are mistakenly placed in the shared queue, \sys can employ \es for tenants in the shared queue to achieve moderate work-conserving bandwidth guarantees in the spirit of \es. Finally, we perform testbed experiments (\S \ref{sec:evaluation:WCBG}) to quantify effects of the binding mechanism: \first the average utilization deficit caused by binding errors is less than $9\%$ of the total capacity; \second to achieve good performance, it is sufficient to perform dynamic binding at more coarse time granularity (\eg a few seconds) compared with the traffic matrix estimation performed at the granularity of milliseconds in~\cite{elasticswitch,trinity}, which would significantly reduce the stress for large scale deployment in practice. 

We plot the system architecture in Figure~\ref{fig:sys_arch}. Next we briefly discuss the components of each module, and defer their design details in \S\ref{sec:tenant_placement} and \S\ref{sec:queue_binding}, respectively.

\subsubsection{Balanced Tenant Placement Module}
The balanced tenant placement module has two components. In particular, given a tenant embedding request, the routing explorer (\S\ref{sec:placement:search}) outputs all Tenant Routing (TR) candidates that can accommodate the tenant. The placement algorithm (\S\ref{sec:option_evaluation}) evaluates each candidate to select the most desired one.

\parab{Tenant Routing (TR) Exploration.} 
As explained in \S \ref{subsec:background}, a tenant $\mathbf{T}$'s TR is a tree in the physical network topology that connects the servers/hypervisors hosting $\mathbf{T}$'s VMs (``servers'' and ``hypervisors'' are used interchangeably). Traffic generated by $\mathbf{T}$'s VMs is confined within its TR. Thus, the TR needs to be provisioned with sufficient VM slots and network bandwidth to fulfill $\mathbf{T}$'s requirement. TR exploration, essentially, is the topology search process that produces a set of virtual networks (\ie overlay trees) that can accommodate $\mathbf{T}$. Admission rules are applied here to accept new tenants only if the datacenter has sufficient spare capacities. 

\parab{TR Candidate Election.} Each TR candidate is evaluated based on two criteria: bandwidth reservation cost and queue occupation cost. Reducing bandwidth reservation cost allows datacenters to accommodate more tenants, while the key reason for considering the queue occupation cost is to reduce the management stress for the dynamic tenant-queue binding module.

\subsubsection{Dynamic Tenant-queue Binding Module}
The dynamic binding module executes \emph{periodically} to distribute queues among tenants. It is built on \first tenant demand trend prediction (\S\ref{sec:queue_binding:demand}), \second the queue-to-tenant allocation algorithm (\S\ref{sec:queue_binding:allocation}) and \third the policy enforcer enforcing allocation decisions inside the network (\S\ref{sec:wfq_configuration}).

\parab{Traffic Demand Trend Prediction.}
Based on the usage measurement in current \emph{control interval}, \sys predicts demand trends of tenants in the next interval, \ie whether a tenant tends to have higher demand than its bandwidth guarantees in the next interval. Note that \sys's prediction relies on only a scalar metric rather than the per-VM pair traffic matrix as proposed in~\cite{elasticswitch,trinity}.

\parab{Queue-to-tenant Allocation.} The allocation algorithm dynamically distributes queues among tenants. In case of queue scarcity, it ranks the competing tenants based on both their demands and payment factors. Considering payment factors mitigates the problem of demand lying by tenants.

\parab{Policy Enforcer.} To enforce bandwidth allocation decisions inside the network, the policy enforcer need to perform a set of tasks include performing network-related operations (\eg switch configuration), tagging tenant packets with proper \textsf{dscp} values and running \es-like rate allocations at hypervisors for tenants without dedicated queues.

%% file: tenant_placement.tex
\newcommand{\algrule}[1][.2pt]{\par\vskip.5\baselineskip\hrule height #1\par\vskip.5\baselineskip}
\section{Balanced Tenant Placement}\label{sec:tenant_placement}
The goals of tenant placement are \first provisioning virtual networks for tenants to satisfy their computation and bandwidth guarantees and \second balancing the overall switch queue utilization among tenants. The prior placement algorithms proposed in \cite{tag, oktopus} aim to maximize the number of accepted tenant requests, which is an NP-hard problem similar to \cite{hose}. However, different from prior algorithms that make greedy embedding decisions (\ie embed a tenant immediately once a feasible option is found), our balanced tenant placement requires \emph{global topology investigation}, \ie evaluating all feasible options before making embedding decisions. Towards this end, we design our own tenant placement algorithm. Formulated in Algorithm~\ref{alg:placement}, our placement algorithm contains two major parts \first TR candidate exploration and \second TR candidate election.

\subsection{TR Candidate Exploration}\label{sec:placement:search}
We first explain TR candidate exploration in the widely adopted multi-rooted tree datacenter topology~\cite{fattree,vl2,tag,silo,jupiter}. Then, we discuss extending such exploration to support randomly connected topology~\cite{jellyfish,small_dc}.

\begin{algorithm}[t]
	\setstretch{0.85}
	\SetKwData{IP}{\textbf{Input:}}
	\SetKwData{OP}{\textbf{Output:}}
	\SetKwData{main}{\textbf{Main:}}
	\SetKwData{return}{\textbf{return}}
	\SetKwData{FC}{\textbf{Function:}}
	\SetKwData{GFTR}{get\_feasible\_TRs\_at\_layer}
	\SetKwData{GTR}{get\_TRs\_at\_layer}
	\SetKwData{GN}{get\_nodes\_at\_layer}
	\SetKwData{STs}{\texttt{trees}}
	\SetKwData{GST}{get\_tree\_for\_root}
	\SetKwData{add}{add}
	\SetKwData{IFF}{is\_feasible}
	\SetKwData{AVM}{available\_VM\_slots}
	\SetKwData{root}{root}
	\SetKwData{leaves}{leaves}
	\SetKwData{GSBC}{get\_tree\_capacity}
	\SetKwData{ETR}{evaluate\_TR}
	\SetKwData{TRC}{TR\_candidates}
	\SetKwData{GDTR}{get\_desired\_TR}
	\SetKwData{GOA}{get\_optimal\_allocation}
	\SetKwData{OA}{optimal\_allocation}
	\SetKwData{GTWLC}{get\_TRs\_with\_least\_BC}
	\SetKwData{GOT}{get\_TR\_with\_least\_QC}
	\SetKwData{Main}{\textbf{Main Procedure:}}
	\small
	
	\IP A tenant request with explicit guarantees. \\
	
	\OP The desired TR or an embedding error.\\
	
	\algrule[0.5pt]
	
	\Main \\
	\Begin{
		$layer \leftarrow 1$\; 
		\While{True}{
			$TRs \leftarrow \GTR{layer}$\; \label{line:getTR}
			\For{$T \in TRs$}{
				$[feasible, cost] \leftarrow \ETR{T}$\;\label{line:iterateTR}
				\lIf{feasible}{$TR\_candidates$.\add{(T, cost)}}
			}
			
			\eIf{$TR\_candidates$ is empty}{ \label{line:nextLayer}
				$layer \leftarrow layer + 1$\; 
				\lIf{$layer > n$}{\return False}\label{line:error}
			}{
				\return \GDTR{$TR\_candidates$}\label{line:election}
			}
		}
	}
	
	\algrule[0.5pt]
	\FC \ETR{$T$}: \\
	\quad $OA \leftarrow \GOA(T)$\; \label{line:optimal_allocation}
	\quad \lIf{OA is feasible}{\return $[True, (c_b, c_q)]$}\label{line:cost}
	\quad \lElse{\return $[False, null]$}
	
	\caption{\bf Balanced Tenant Placement}\label{alg:placement}
\end{algorithm}
\normalsize

Given a tenant request, Algorithm \ref{alg:placement} explores the topology from the lowest layer (hypervisor layer) towards the highest layer (core switch layer). At each layer, function \textsf{get\_TRs\_at\_layer} (line \ref{line:getTR}) obtains all TR options at this layer. A layer-$i$ TR option is a tree rooted at layer $i$. Its leaves are the servers reachable from the root using only downward paths. Then the algorithm evaluates these TRs  to produce \emph{feasible} ones, called TR candidates (line \ref{line:iterateTR}). Generally speaking, a TR option is feasible if it has enough capacity to accommodate the tenant. Function \textsf{evaluate\_TR}, detailed in \S \ref{sec:option_evaluation}, determines such feasibility. If no TR candidates can be found, the algorithm continues exploration in the next layer (line \ref{line:nextLayer}). Otherwise, it stops further exploration and returns the desired TR elected from all candidates (line \ref{line:election}) using the criteria described in \S \ref{sec:option_evaluation}. The early return confines tenants at the lowest possible layer to avoid unnecessary network usage at higher layers. If no TR candidates can be found after exploring the entire topology with $n$ layers, the algorithm returns false (line \ref{line:error}), indicating an embedding error due to the lack of resources.

\parab{Random Topology.} To support random topology, Algorithm \ref{alg:placement} adopts the $k$-shortest path algorithm~\cite{shortest} to obtain a set of paths between each hypervisor pair and then combines  them to produce TR options. The parameter $k$, similar to $layer$ in Algorithm~\ref{alg:placement}, 
determines TR exploration space.

\subsection{TR Evaluation and Candidate Election}\label{sec:option_evaluation}
\begin{figure}[t]
	\centering
	\mbox{
		\subfigure[Requring $\mathbf{B}$ on $S_1 {\leftrightarrow} H_1$
		\label{fig:placement:a}]{\includegraphics[scale=0.4]{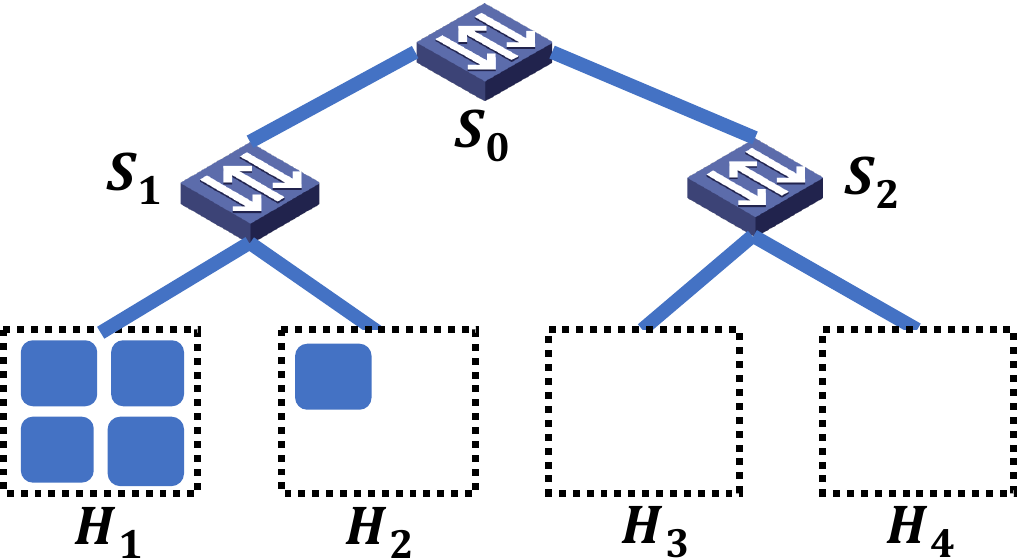}}~
		\subfigure[Requring $2\mathbf{B}$ on $S_1 {\leftrightarrow} H_1$
		\label{fig:placement:b}]{\includegraphics[scale=0.4]{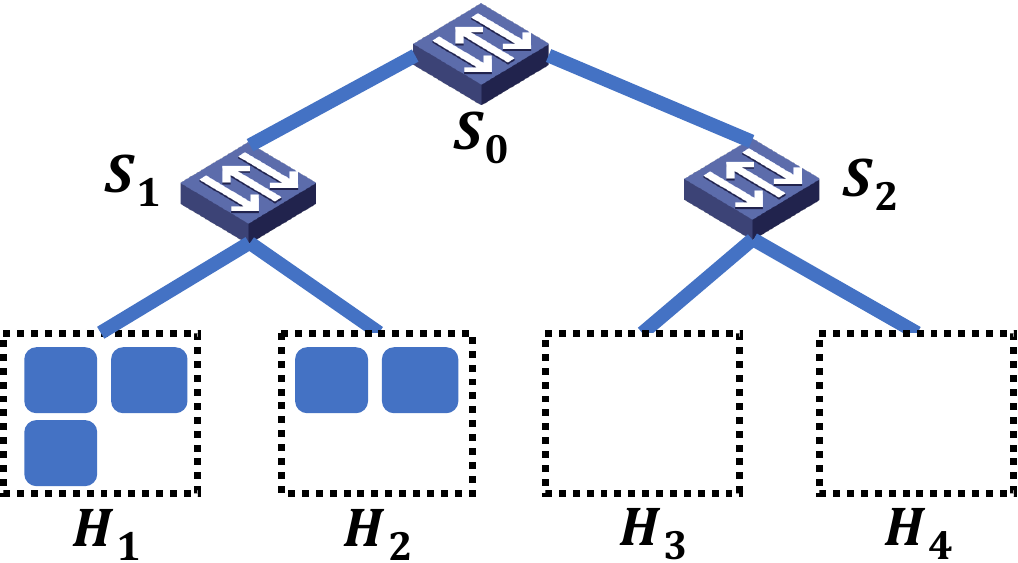}}
	}
	\caption{Given the TR rooted at $S_1$, the bandwidth required on its link depends on VM locations inside the TR.}\label{fig:placement}
\end{figure}

A TR option is feasible if \first the total available VM slots from all its servers are enough to hold the tenant's VMs and \second each link of the TR has enough available capacity to satisfy the tenant's bandwidth guarantees.  Although evaluating the first rule is straightforward, the second rule requires more investigation. In particular, given a TR option, the amounts of bandwidth required on its links depend on the VM locations inside the TR. Specifically, consider a homogeneous hose model where all VMs have the same inbound and outbound bandwidth guarantee $\mathbf{B}$. Given a link $L$ of the TR, removing $L$ breaks the TR into two disjoint components. If $m$ VMs are in one component and $n$ VMs are in the other one, then the  bandwidth required on $L$ is $\mathbf{B}\cdot\min\{m,n\}$. Figure~\ref{fig:placement} plots a TR rooted at $S_1$. For the VM location in Figure~\ref{fig:placement:a},  the two links ($S_1 {\leftrightarrow} H_1$ and $S_1 {\leftrightarrow} H_2$) both need to reserve $\mathbf{B}$ ($\min\{\mathbf{B}, 4\cdot\mathbf{B}\}$) bandwidth whereas they have to reserve $2\mathbf{B}$ for the VM location in Figure~\ref{fig:placement:b}.

To reduce the total network bandwidth required for embedding the tenant, function \textsf{get\_optimal\_allocation} (line \ref{line:optimal_allocation}) produces the VM location that requires the least bandwidth reservation. For homogeneous hose models, the optimal allocation is produced as follows: \first find the server $H$ in the TR with the largest \emph{usable VM slots}, \second allocate as many VMs as possible to $H$, \third  update the remaining network/server 
capacity after allocation and \forth repeat step one until either all VMs are allocated (indicating feasibility) or all servers in the TR have been investigated (indicating infeasibility). The usable VM slots for $H$ in the TR is restricted by both the available VM slots in $H$ and the available bandwidth on the path from $H$ to the TR's root. For instance, in Figure \ref{fig:placement}, if we assume the available bandwidth on link $S_1 {\leftrightarrow} H_1$ is less than $\mathbf{B}$, the usable VM slots in $H_1$ is $0$, rather than $4$.

If the TR's optimal allocation is feasible, it becomes a candidate for embedding the tenant. Algorithm \ref{alg:placement} then computes its bandwidth cost $c_b$ as the sum of reserved bandwidth for the tenant on each link of the TR, and the  queue occupation cost $c_q$ as the largest number of tenants served by any of the TR's links (line \ref{line:cost}). 

Candidate election is based on both $c_b$ and $c_q$. Each TR candidate is associated with a $cost$ combining $c_b$ and $c_q$. The desired TR is the one with lowest  $cost$. One strategy for computing   $cost$ is assigning more weight to $c_q$ when the datacenter load is light to prefer more balanced placement whereas assigning more weight to $c_b$ for heavy-loaded network to prefer the placement with fewer bandwidth cost.

\parab{Supporting High Availability.} Algorithm~\ref{alg:placement} can be extended to support high availability~\cite{high_ava}. In \cite{high_ava}, the worst-case survival ratio (WCS) is defined as the smallest fraction of VMs remaining functional during a single point failure. Consider server as the fault domain. Given a tenant with $N$ VMs and WCS as $f$, one server can host at most $(1{-}f)N$ VMs for this tenant.  By patching the constraint in function \textsf{get\_optimal\_allocation}, Algorithm~\ref{alg:placement} can produce TR that satisfies the high availability requirement. 

\parab{Search Complexity.} 
The search complexity for embedding tenants depends on the layers at which Algorithm~\ref{alg:placement} returns. In a fattree topology~\cite{fattree}, the worse case complexity (\ie the algorithm returns at the core switch layer) is $O(V^{\frac{5}{3}})$, where $V$ is the number of nodes in the network. Thus, although our algorithm perform comprehensive topology search, its time complexity is polynomial rather than exponential. For topologies built with higher over-subscription ratios than fattree, the search complexity is smaller as the number of TR options at each layer is smaller. Further, the topology search results can be cached to achieve long-term efficiency~\cite{OpReduce}. 

%% file: queue_binding.tex
\section{Dynamic Tenant-Queue Binding}\label{sec:queue_binding}
To support more tenants with limited number of queues, \sys's design is inspired by how the working set of a process is often much smaller than the total memory it consumes. Similarly, only tenants whose traffic demands exceed their bandwidth guarantees need dedicated queues. Thus, there is an opportunity for \sys to dynamically allocate limited number of queues to high-demanded tenants. In particular, \sys periodically evaluates each tenant and allocates queues 
among tenants based on their \emph{scores}. Each tenant's score encapsulates its \emph{usage factor} (\S \ref{sec:queue_binding:demand}) and \emph{payment factor} (\S \ref{sec:tenant_score_computation}) so as to prioritize high-demanded and honest tenants.

\subsection{Tenant Demand Trend Prediction}\label{sec:queue_binding:demand}
Because prior works~\cite{elasticswitch,trinity} rely on Guarantee Partitioning (GP) to achieve bandwidth guarantees, they need to predict each tenant's traffic matrix, \ie per VM-pair traffic demand. However, since tenant applications are often agnostic to cloud operators, it is challenging in practice to capture real-time communication patterns among VMs and predict traffic demand between each VM pair. Realizing that, \sys's tenant-queue binding module only relies on predicting whether a tenant tends to have higher demands than its guaranteed bandwidth. Thus, rather than predicting traffic matrix, \sys proposes to use a scalar metric, usage factor (U-factor), to indicate a tenant's network utilization with respect to its guaranteed bandwidth. We do not claim that U-factor is the optimal metric for demand prediction. However, it does greatly reduce the stress of prediction by focusing on tenant-level \emph{demand trend} rather than VM-level traffic matrix.

Each tenant's U-factor is computed per \emph{control interval}. Specifically, in each control interval, all hypervisors measure the bandwidth utilization of their hosted VMs. As VMs can have both inbound and outbound traffic, bi-directional bandwidth usage is considered. For instance, consider a hypervisor $H_j$ hosting $m$ VMs of a tenant $\mathbf{T}$.  Then $\mathbf{T}$'s inbound (outbound) bandwidth usage $U_j^{i}$ ($U_j^{o}$) measured by $H_j$ is the sum of inbound (outbound) bandwidth usage from all these $m$ VMs.  

At the end of each control interval, \sys computes each tenant's U-factor. For tenant $\mathbf{T}$, one way of computing its U-factor $\mathbf{U}_\mathbf{T}$ is as follows
\begin{equation}\label{equ:u_factor}
\mathbf{U}_\mathbf{T} = \min\{\max\limits_{H_j \in \mathbf{H}} \frac{\max\{U_j^{i}, U_j^{o}\}}{B_j}, 1\}, 
\end{equation}
where $\mathbf{H}$ is the set of hypervisors hosting $\mathbf{T}$'s VMs and $B_j$ is $\mathbf{T}$'s guaranteed bandwidth on $H_j$'s network interface. 
If $H_j$ hosts $m$ VMs from $\mathbf{T}$ (provisioned with total $N$ VMs), $B_j{=}\mathbf{B}\cdot\min\{m,N{-}m\}$ considering a symmetric and homogeneous model with per-VM guarantee $\mathbf{B}$.

The design rationale of Equation~(\ref{equ:u_factor}) is as follows. The innermost \textsf{max} is necessary as  the high-demanded VMs may either send or receive large volumes of traffic. The middle \textsf{max} is designed to handle  many-to-one traffic pattern in which many source VMs in remote servers are communicating with a few destination VMs in the local server. Although source hypervisors may measure small usage since source VMs are bottlenecked by destination VMs. $\mathbf{T}$ actually has large traffic demand at these receivers. Taking the largest usage among all hypervisors will capture such a communication pattern. Finally, the outermost \textsf{min} sets a  $\mathbf{U}_\mathbf{T}$ cap of $1$. We leave the exploration of other U-factor definitions  in future work.

\subsection{Tenant Intent Lying Mitigation}\label{sec:tenant_score_computation}
Merely using U-factors to allocate queues has problems when tenants lie about their real bandwidth guarantees: a tenant can deliberately request smaller guaranteed bandwidth so as to have high U-factors. Note that no work-conserving allocation policies can completely prevent tenants from gaining advantages via lying, \ie being strategy-proof~\cite{faircloud}. To mitigate the problem caused by lying, \sys proposes to consider payment factors, along with U-factors, when scoring tenants. Each tenant's payment factor and its guaranteed bandwidth are positively correlated such that deliberately requesting lower guarantees reduces a tenant's score whereas exaggerating guarantees requires higher payment. As designing pricing model is not the focus of this paper, \sys assumes, for simplicity, that a tenant's payment factor is proportional to the total guaranteed bandwidth required by its hose model\footnote{Payment for computation resources is not considered as \sys focuses on network bandwidth management.}. Thus, given tenant $\mathbf{T}$ with $N$ VMs and each VM requests guaranteed bandwidth $\mathbf{B}$, its payment factor is $kN\mathbf{B}$, where $k$ is a constant depending on the pricing model. 

Simplifying $\mathbf{U}_\mathbf{T}$ in Equation (\ref{equ:u_factor}) as $\min\{ \frac{U^*}{B^*}, 1\}$, then tenant $\mathbf{T}$'s score $S_\mathbf{T}$ is computed as follows 
\begin{equation}\label{equ:score}
S_\mathbf{T} = kN\mathbf{B} \cdot \mathbf{U}_\mathbf{T} = 
\begin{cases}
\tilde{k}U^*, & \text{if}\ \mathbf{U}_\mathbf{T} < 1 \\
kN\mathbf{B}, & \text{otherwise}
\end{cases}
\end{equation}
$\tilde{k}=kN\mathbf{B}/B^*$, where $B^*$ is determined by maximizing the inner \textsf{max} operation of Equation~(\ref{equ:u_factor}).

Using $S_\mathbf{T}$ as the criterion for queue allocation can mitigate problems caused by lying. On the one hand, as $S_\mathbf{T}$ is bounded by $kN\mathbf{B}$, deliberately requesting smaller $\mathbf{B}$ would result in a lower cap of $S_\mathbf{T}$, which is disadvantageous when competing with other tenants. On the other hand, deliberately requesting higher $\mathbf{B}$ than real demand also has problems as \first tenant $\mathbf{T}$ would have to pay more and \second its $S_\mathbf{T}$ would be determined by $\mathbf{T}$'s real usage rather than its claimed guarantees if $\mathbf{T}$ has smaller demands than its guarantees (\ie $\mathbf{U}_\mathbf{T}<1$). Generally, tenants with higher traffic demands are preferred since $S_\mathbf{T}$ is non-decreasing as bandwidth usage increases. Such a property is desired for the queue allocation detailed below.

\subsection{Dynamic Queue Allocation}\label{sec:queue_binding:allocation}
We present the queue allocation logic in Algorithm~\ref{alg:allocation}. A tenant is assigned a dedicated queue only if it is assigned a dedicated queue on each link of its TR. Otherwise, the tenant will be served in the shared queue on each link of its TR. To prioritize tenants with higher scores, Algorithm~\ref{alg:allocation} starts queue assignment from the tenant with the highest score, breaking tie randomly (line \ref{line:score}). 

If a tenant $T$ already occupies a dedicated queue, it continues to hold the queue (line~\ref{line:already}) for the next control interval. This indicates that $T$ either maintains its high score or owns a dedicated queue on each link of its TR due to the lack of queue contention, which is possible due to the balanced placement (see analysis in \S\ref{sec:evaluation:queue_scarcity}).

If tenant $T$ is currently placed in the shared queue, Algorithm \ref{alg:allocation} determines whether allocating $T$ a dedicated queue is possible. To satisfy the condition on line~\ref{line:spare}, each link of $T$'s TR needs to have at least one spare queue. If positive, function \textsf{enqueue\_tenant} assigns $T$ a queue on each link $L$ of its TR (line \ref{line:enqueue}).

\begin{algorithm}[t]
\setstretch{0.85}
\SetKwData{IP}{\textbf{Input:}}
\SetKwData{OP}{\textbf{Output:}}
\SetKwData{main}{\textbf{Main:}}
\SetKwData{return}{\textbf{return}}
\SetKwData{FC}{\textbf{Function:}}
\SetKw{TT}{\texttt{tenant}}
\SetKwData{GBWC}{get\_optimal\_placement}
\SetKwData{GTWLC}{get\_TRs\_with\_least\_BWC}
\SetKwData{EQ}{enqueue\_tenant}
\SetKwData{OEN}{opportunistically\_enqueue}
\SetKwData{GOQ}{get\_opportunistic\_queues\_from\_LSTs}
\SetKwData{continue}{\textbf{continue}}
\SetKw{=}{$\leftarrow$}
\small

\IP The set of embedded tenants $\mathcal{S}$. \\
\OP Tenant-queue assignment. \\
\algrule[0.5pt]
Sort the tenants in $\mathcal{S}$ decreasingly by their scores; \\ \label{line:score}
\For{T $\in \mathcal{S}$}{
	\lIf{T has a dedicated queue}{\continue} \label{line:already}
	\ElseIf{T's TR has a spare queue}{ \label{line:spare}
		\EQ{T}\;
	}
	\lElse{\OEN{T}}\label{line:find_op}
	Update queue allocation state\;\label{line:update}
}
Queue weight computation\; \label{line:queue_weight}

\algrule[0.5pt]

\FC \EQ{T}: \\ 
\For{L $\in$ T's TR}{\label{line:enqueue}
	$reserved\_bandwidth$ \= $\mathbf{B}\cdot\min\{m,N-m\}$\;
}

\algrule[0.5pt]

\FC \OEN{T}: \\ 
\For{L $\in$ T's TR}{
	\GOQ{L}\; \label{line:op_enqueue}
}
\If{T's TR has an opportunistic queue}{\label{line:has_op}
	\EQ{T}\;}
\caption{\bf Queue Allocation Algorithm}\label{alg:allocation}
\end{algorithm}
\normalsize

Finally, if at least one link of $T$'s TR runs out of queues, function \textsf{opportunistically\_enqueue} (line \ref{line:find_op}) \emph{opportunistically} finds queues for $T$ by preempting queues from low-scored tenants (LSTs). Specifically, on link $L$ without spare queues,  the algorithm obtains an opportunistic queue occupied by a tenant $\tilde{T}$ such that \first $\tilde{T}$'s score is less than $T$'s score and \second $\tilde{T}$'s score is the smallest among all tenants owning a queue on $L$ (line \ref{line:op_enqueue}). If an available queue, either opportunistic or unoccupied, exists on each link of $T$'s TR, we say that $T$'s TR has an opportunistic queue (line \ref{line:has_op}), and then enqueue $T$. The dequeued tenants will be served in shared queues during the next control interval. 

Queue allocation state is updated after handling $T$ (line~\ref{line:update}). Once queue allocations for all tenants are finished, \sys computes weight for each queue (line~\ref{line:queue_weight}). For a queue $\mathcal{Q}_i$ on link $L$, its normalized weight is the ratio of reserved bandwidth in $\mathcal{Q}_i$ to the total reserved bandwidth on link $L$. In practice, weights need to be proportionally translated into the supported values on commodity switches (\eg $1$ to $15$ on our switches).

Tenant departures will trigger network allocation state update as well. Newly arrived tenants are served in shared queues, and will be evaluated at the end of current interval.

\subsection{Policy Enforcer}\label{sec:wfq_configuration}
To enforce queue allocation decisions inside the network, \sys needs to perform \first packet tagging and \second network configuration. Packet tagging is to ensure that packets from different tenants can be correctly identified and therefore served in correct queues. We use \textsf{dscp} tagging to achieve this. To avoid ambiguity, the D-tenants (tenants with dedicated queues) whose TRs share at least one common link cannot use the same \textsf{dscp} value. D-tenants whose TRs are non-overlapping can reuse the same \textsf{dscp} value. Given that \textsf{dscp} values range from $0$ to $63$, finding the smallest possible number of \textsf{dscp} values in a legal assignment can be reduced to the k-coloring of a graph, which is NP-hard~\cite{graph_coloring}. 

To address the \textsf{dscp} usage concern, we analyze the efficiency of a greedy assignment in large scale datacenters based on production datacenter settings. The results show that $64$ \textsf{dscp} values are sufficient to avoid conflict even when the datacenter is fully reserved (see details in \S\ref{sec:evaluation:queue_scarcity}). Further, technically, it is possible to mutate \textsf{dscp} values on switch ports via the \textsf{dscp}-to-\textsf{dscp} mutation map~\cite{dscp_swap}. Thus, based on a \textsf{dscp} mapping on each port, a tenant can use different \textsf{dscp} values on different ports, eliminating the static \textsf{dscp} value reservation required on each link of its TR, which in turn eliminates the possibility of \textsf{dscp} conflict.

Network configuration involves configuring queues on each  link with proper weights and \textsf{dscp} values, which requires WFQ configuration on both ports of the link. For edge links connecting servers and switches, software WFQ is required on hypervisors. To support automation, \sys designs a network action container to perform configuration in a batch:  operations on different switches are parallelized via multi-threading so that the marginal configuration latency is negligible.

The final part of the policy enforcer is that \sys can run \es-like rate allocation mechanisms~\cite{seawall,seawall-like} for tenants without dedicated queues to provide them bandwidth guarantees and achieve moderate work conservation in the worse case when all D-tenants have insufficient demands. However, \sys imposes smaller overhead than \es~\cite{elasticswitch} since it only performs rate allocations for tenants without dedicated queues.

%% file: implementation.tex
\section{Implementation}\label{sec:implementation}
\begin{figure}[t]
  \centering
  \mbox{
	\subfigure{\includegraphics[scale=0.44]{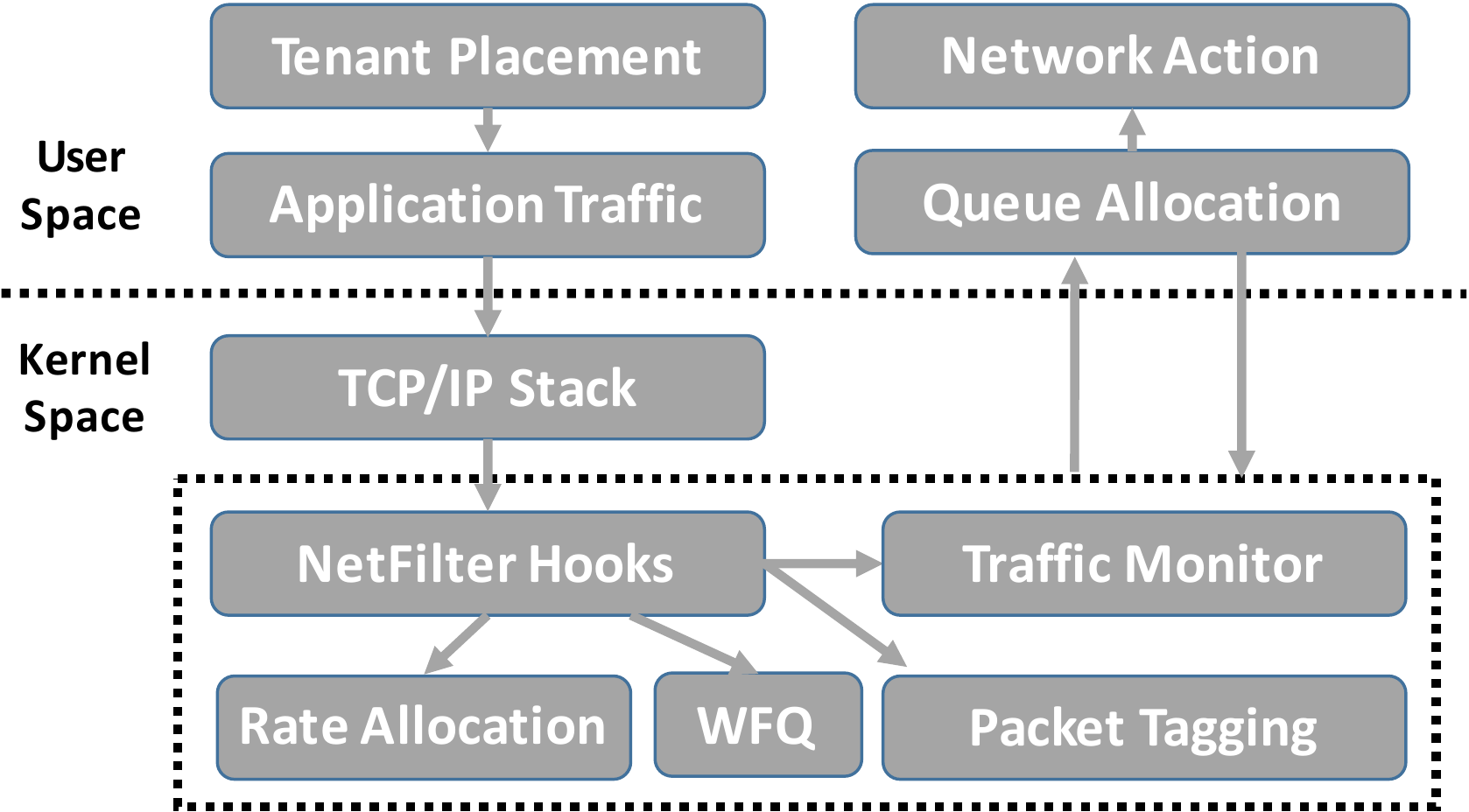}}
  }
  \caption{The software implementation of \sys.}
\label{fig:software_stack}
\vspace{0.02in}
\end{figure}

In this section, we introduce the prototype implementation of \sys. The prototype of \sys contains both user-space and kernel-space programs, as plotted in  Figure~\ref{fig:software_stack}. The user-space programs, executed globally, are responsible for managing the whole datacenter whereas the kernel-space program, running on each hypervisor, manages the local hypervisor. Two spaces interact with each other such that queue allocation decisions are made based on the distributed measurement reported by all hypervisors, and afterwards the allocation decisions are pushed back to the kernel module for enforcement on each hypervisor. The current implementation has ${\sim}2000$ lines of code (\textsf{Python} in user space and $\textsf{C}$ in kernel space). 

The user-space programs include tenant placement, queue allocation and the network action container. The kernel-space module, built on \textsf{NetFilter}~\cite{netfilter}, includes tenant traffic monitor, rate allocation  (for tenants in shared queues), software WFQ (for tenants with dedicated queues) and packet \textsf{dscp} tagging. On each hypervisor, a user-space deamon (not plotted) based on \textsf{Netlink}~\cite{netlink} interacts with the 
kernel module. 

Note that implementing a hypervisor that can support all kinds of VM management is out of this paper's scope. Our prototype builds a simple hypervisor that can support \sys-related operations, such as identifying the VMs of each tenant.

%% file: evaluation.tex
\section{Evaluation}\label{sec:evaluation}
\begin{figure*}[t]
  \centering
  \mbox{
	\subfigure[Per-tenant runtime bandwidth utilization given 
	predictable demand trends.\label{fig:5HDT}]{\includegraphics[scale=0.45]{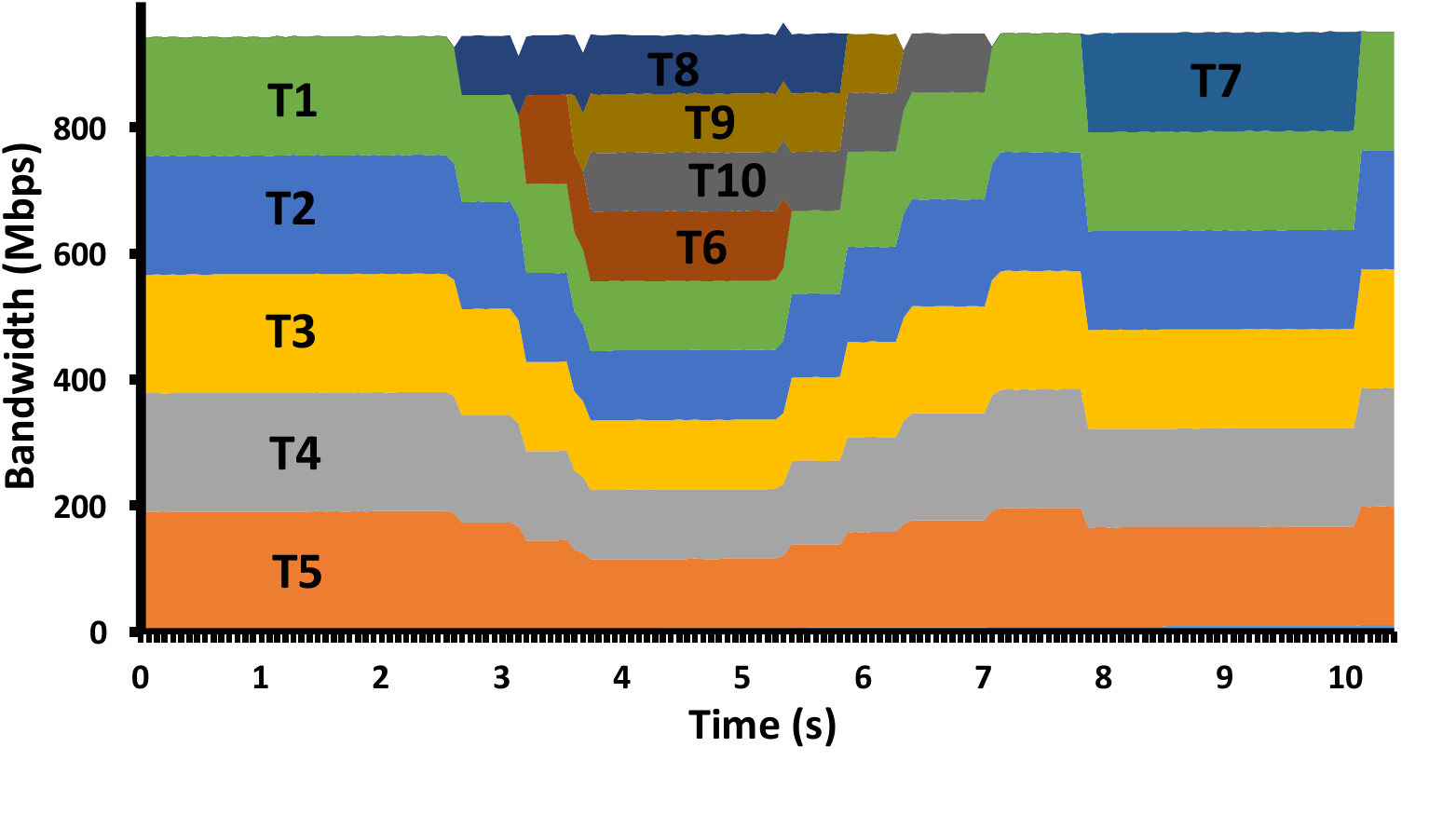}}\quad\quad
	\subfigure[Aggregate core link utilization given 
	unpredictable traffic trend.\label{fig:mapreduce:a}]{\includegraphics[scale=0.27]{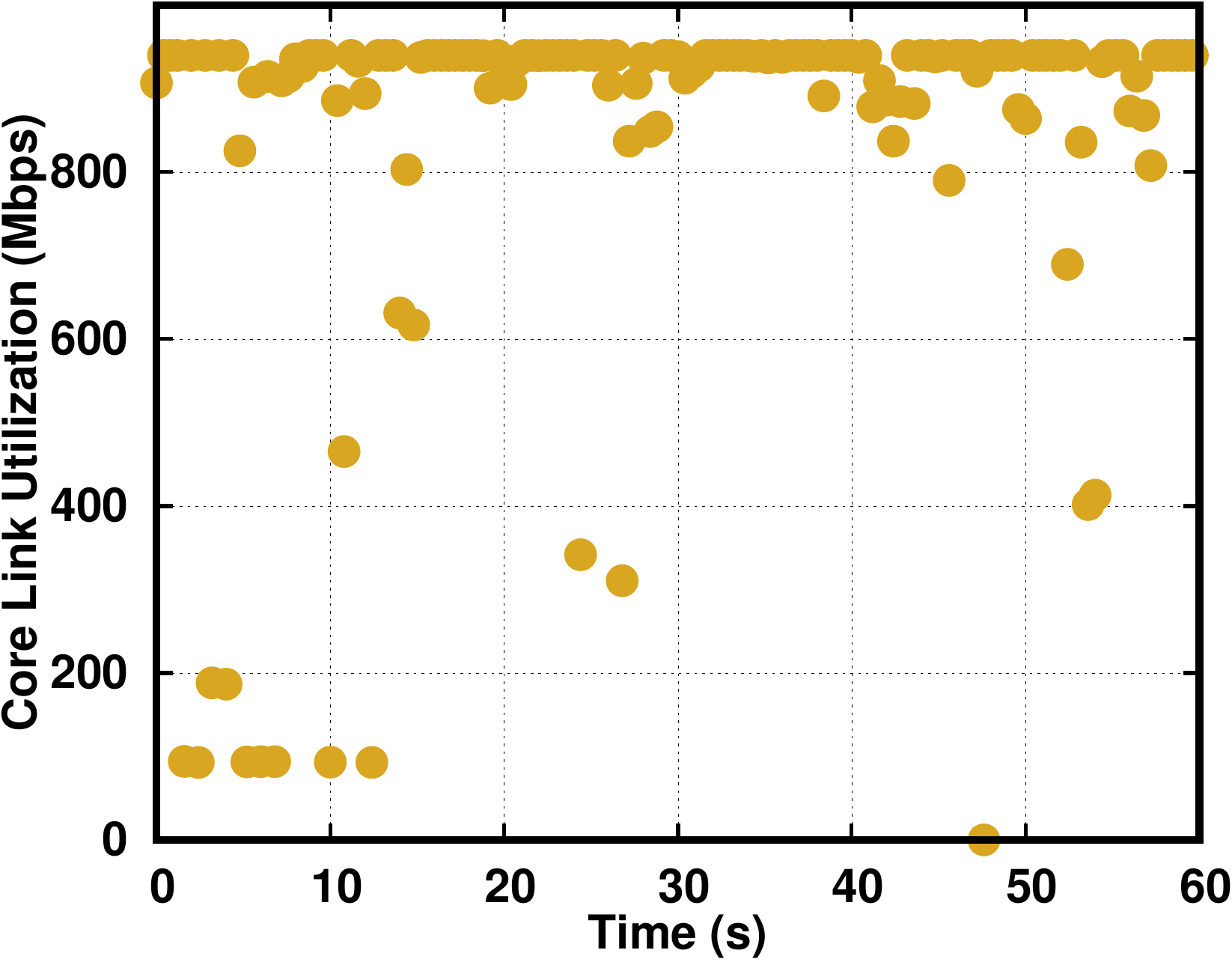}}\quad\quad
	\subfigure[Avg core link utilization with 
	varying control intervals.\label{fig:mapreduce:b}]{\includegraphics[scale=0.27]{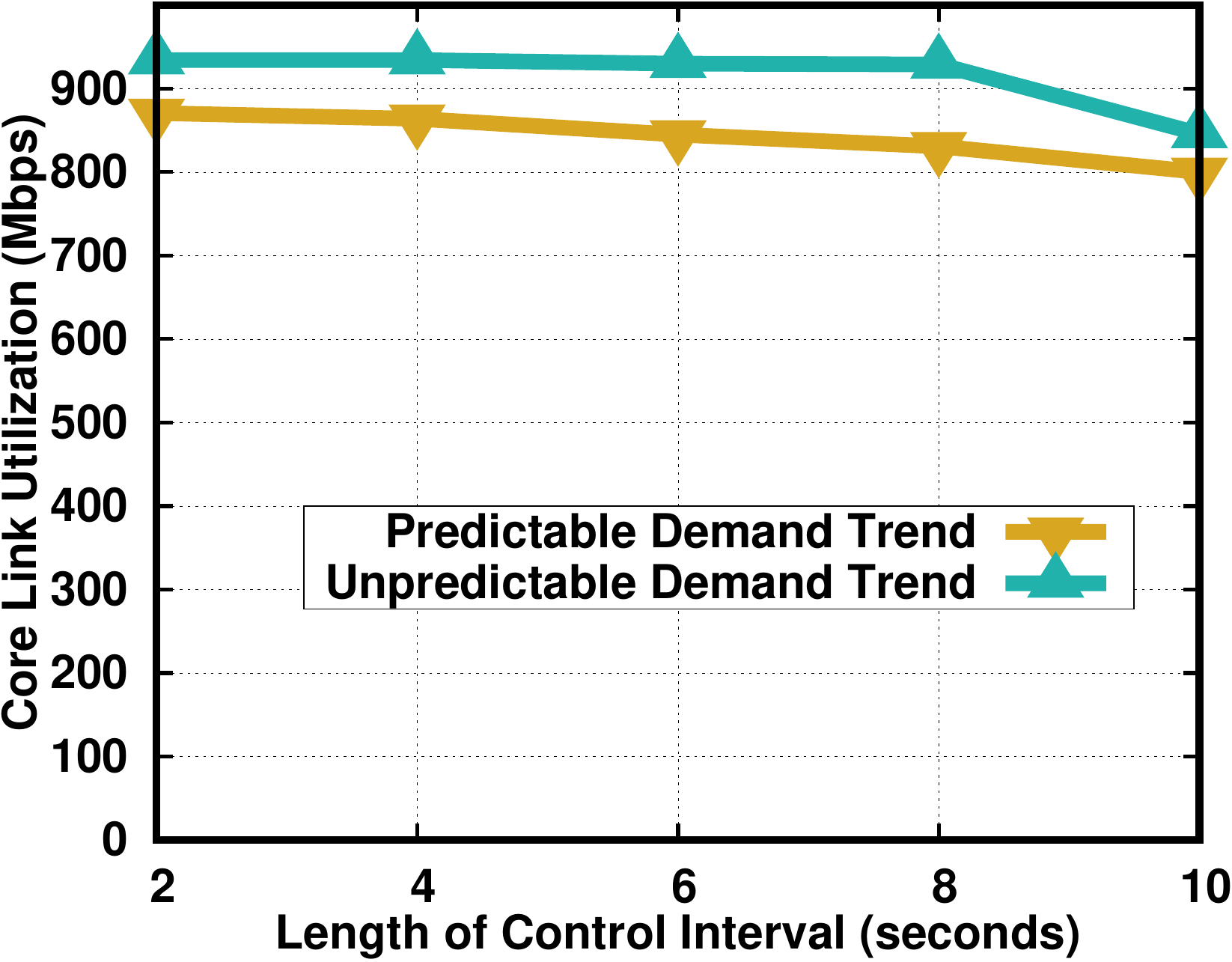}}
  }
  \caption{Figure~\ref{fig:5HDT} plots runtime bandwidth utilization of all tenants given correct demand trend prediction. \sys achieves perfect work-conserving bandwidth guarantees in this case. Figure~\ref{fig:mapreduce:a} plots the total runtime utilization given completely unpredictable demands. Only few under-utilized cases are observed during the measurement period, yielding over $91\%$ average utilization. Figure~\ref{fig:mapreduce:b} shows the average link utilization given different lengths of the control interval.}
\label{fig:mapreduce}
\end{figure*}

Our evaluation centers around the following questions: 

\noindent\textbf{\first How does traffic dynamic affect \sys's performance?} With correct predictions on demand trend (not the exact traffic matrix), \sys achieves \emph{perfect} work-conserving bandwidth guarantees: all bandwidth guarantees are satisfied and meanwhile the bottleneck link is fully utilized (\S\ref{sec:evaluation:predictable}). Even when demand trends are \emph{completely unpredictable}, \sys drives the bottleneck link to over $91\%$ utilization (\S\ref{sec:evaluation:unpredictable}) without comprising bandwidth guarantees.

\noindent\textbf{\second How well can \sys benefit applications?}
Given the above desirable properties, \sys significantly benefits applications, for instance, by reducing their flow completion times (FCTs) by up to $50\%$ compared with the state-of-the-art solutions~\cite{elasticswitch,trinity} (\S\ref{sec:evaluation:FCT}). 

\noindent\textbf{\third How well can \sys manage large scale datacenters?} 
Based on observations from production datacenters, we analyze \sys in a large scale datacenter. We show that \sys can assign dedicated queues to ${\sim}90\%$ of the tenants in any control interval even when the datacenter is fully reserved. Thus, \sys produces at least $3\times$ throughput gain over the guarantees and achieves higher efficiency in  link utilization (\S\ref{sec:evaluation:large_scale}).

\noindent\textbf{\forth How much overhead does \sys impose?} 
\sys imposes small overhead for switch configuration, running rate allocations and embedding tenants (\S\ref{sec:evaluation:benchmark}).

\parab{Testbed Experiment Setup}. We build a physical testbed containing $10$ servers and each server provisions $10$ VM slots, for a total of $100$ VMs. Each server installs a Gigabit Ethernet NIC and runs the $3.13.0$ Linux kernel. We evenly distribute the servers into two racks inter-connected by two Pronto-3297 48-port Gigabit (ToR) switches. Thus, the topology is $5{:}1$ oversubscribed and  the core link may be congested when VMs are sufficient demands. Each port supports up to $8$ WFQ queues. We embed multiple tenants in the testbed, with random sizes from $2$ to $20$ VMs.

We develop a client/server program to generate traffic. The clients initiate long-lived TCP connections to randomly selected servers and request flow transmission. All VMs run both the client and server programs. Only intra-tenant communication is allowed.

\subsection{Work-Conserving Bandwidth Guarantees}\label{sec:evaluation:WCBG}
In this section, we consider how traffic dynamics may affect \sys's performance for enabling work-conserving bandwidth guarantees. We consider the following two scenarios. The first case is that a tenant's demand trend is predictable: \ie once a tenant has high traffic demand, this trend continues for few seconds. Trend predictability is not over-optimistic since hot spots in production datacenters can last over tens of seconds~\cite{dc_mesaure}. The second case is that the demand trend is \emph{completely unpredictable}: \ie a tenant's future demands are independent on its current or previous demands. In both cases, \sys does not impose any constraint on VM communication patterns, \ie one client can request flow transfers from arbitrary servers at any time. 

To quantify the worst-case performance degradation caused by traffic unpredictability, we first disable the \es-like rate allocations for the tenants without dedicated queues, and allocate them at most their guaranteed bandwidth. 

\subsubsection{Predictable Demand Trend}\label{sec:evaluation:predictable}
In this experiment, we consider $10$ tenants competing on the core link. Each tenant is guaranteed $94$Mbps bandwidth on the core link. To generate traffic, we randomly pick $5$ tenants (referred to as T1 to T5) as high-demanded  tenants whose clients request sufficient flow transfers during our measurement period.  The remaining tenants (referred to as T6 to T10) have insufficient demands during the measurement period. Low-demanded tenants may initiate their flow transfers at any time during the measurement period. In this experiment we first fix the length of control interval as $4$ seconds. Different settings are considered in \S\ref{sec:evaluation:unpredictable}.

Figure~\ref{fig:5HDT} plots the runtime core link bandwidth obtained by each tenant in a 10-second measurement period. During this period, \sys's tenant-queue binding algorithm assigns each of the tenants in T1 to T7 a dedicated queue on the core link; T8, T9 and T10 are served in a shared queue. When low-demanded tenants are inactive at the early stage, T1 through T5 fairly share the entire core link capacity. Later on, low-demanded tenants T6, T8, T9 and T10 become active. As T8, T9 and T10 are in the shared queue, they all obtain their guaranteed bandwidth. T1 to T6, each exclusively occupying a queue, equally share the remaining capacity. At about $8$ second, T7 becomes active and fairly shares the core link with T1 to T5. It is clear that all tenants receive at least their guaranteed bandwidth regardless of their communication patterns and other tenants' traffic demands. Meanwhile, the core link is always fully utilized. Thus, \sys achieves perfect work-conserving bandwidth guarantees.

\begin{figure*}[t]
	\centering
	\mbox{
		\subfigure[Overall average FCTs]{\includegraphics[scale=0.27]{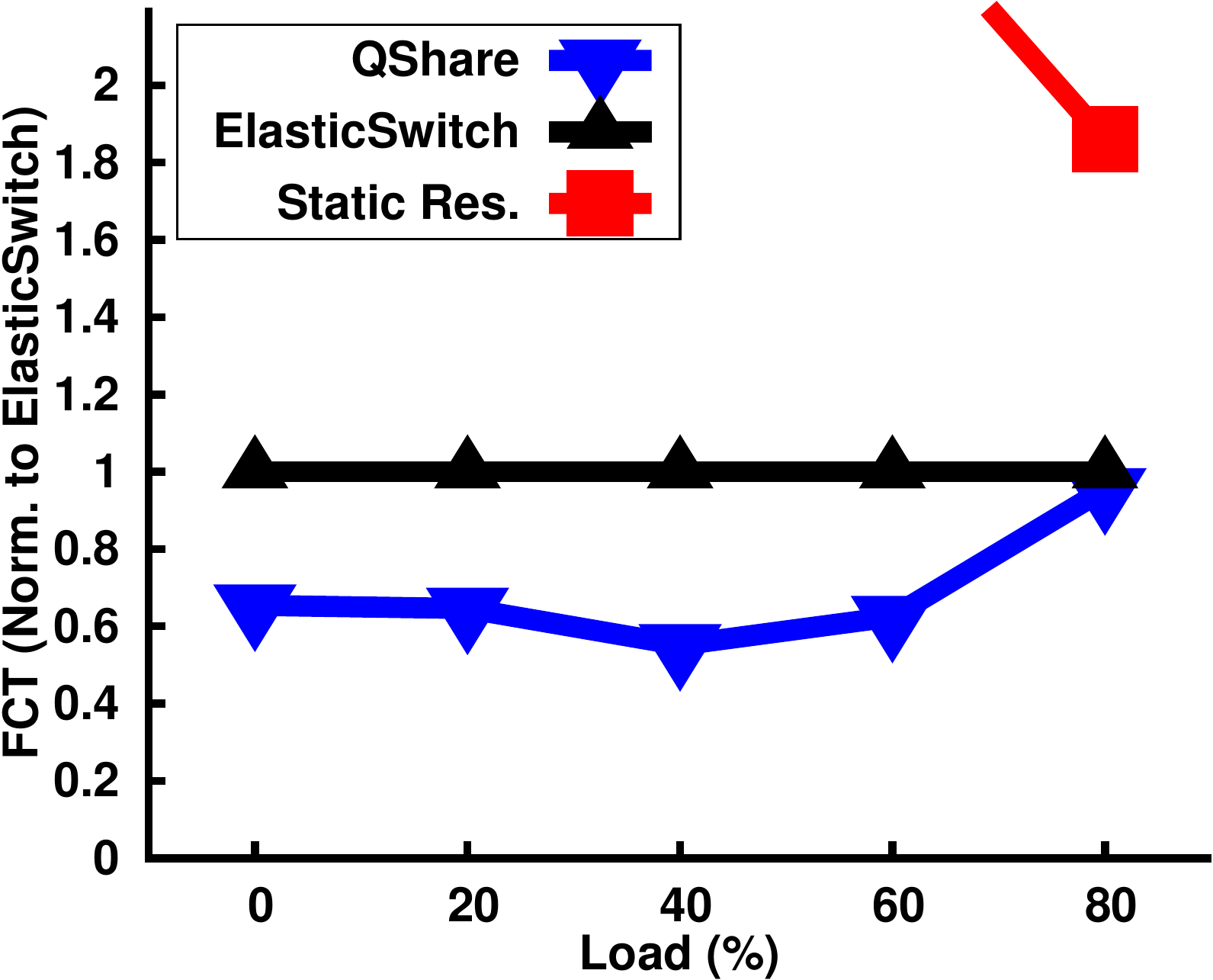}}
		\subfigure[Small flows (${\leq}100$KB)]{\includegraphics[scale=0.27]{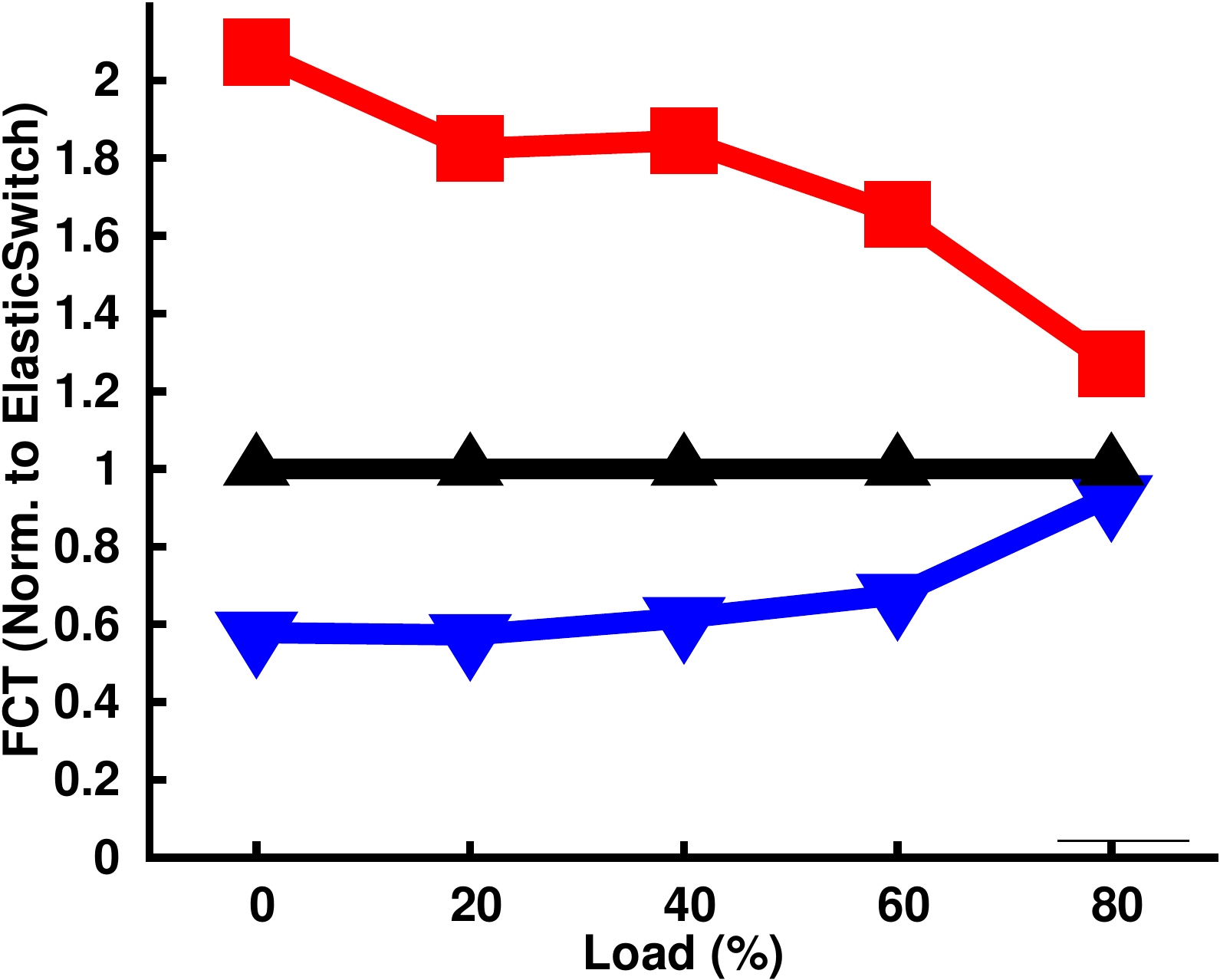}}
		\subfigure[Medium flows]{\includegraphics[scale=0.27]{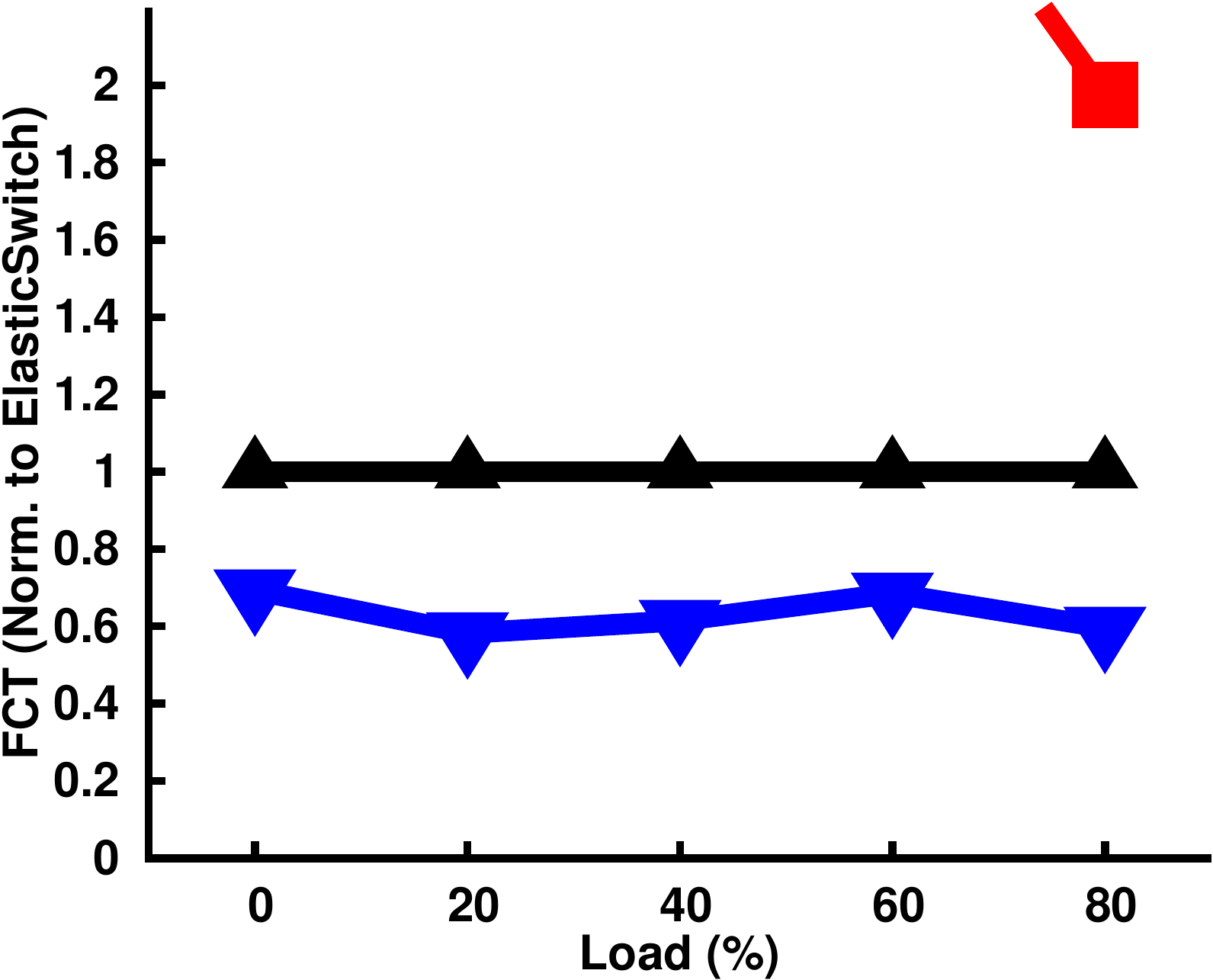}}
		\subfigure[Large flows (${\geq}10$MB)]{\includegraphics[scale=0.27]{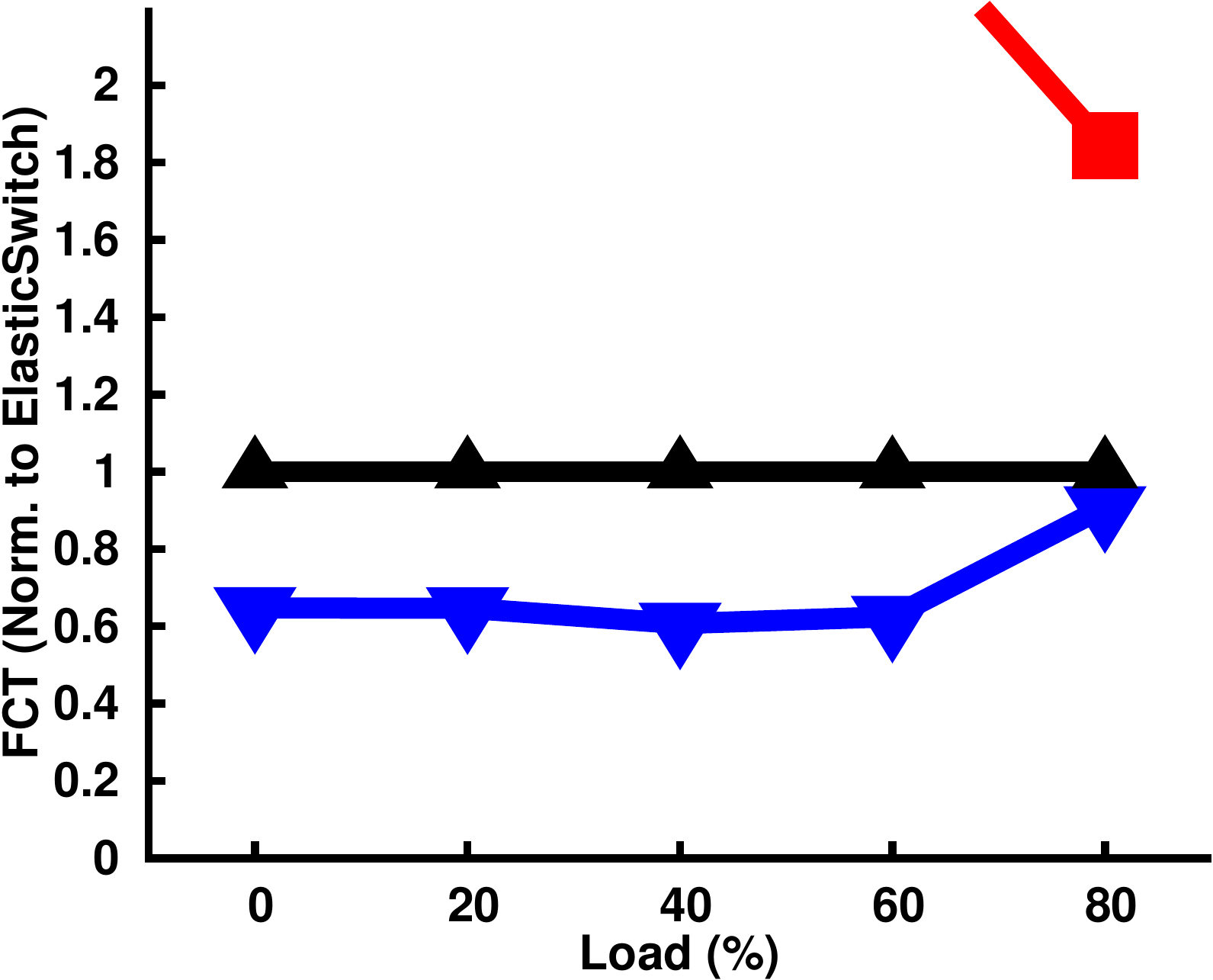}}
	}
	\caption{FCT statics for varying fabric loads (part of the results for static reservation are out of the plot scope). Despite its improvement over static reservation, \es~\cite{elasticswitch} has a large performance degradation compared with \sys (up to $2\times$ long FCTs) even if it adopts aggressive RA at the expense of compromising bandwidth guarantees. In term of bandwidth utilization, Trinity has roughly the same performance as \es with aggressive RA.} 
	\label{fig:fct}
\end{figure*}
 
\subsubsection{Unpredictable Demand Trend}\label{sec:evaluation:unpredictable}
In this section, we consider the case when tenant demand trend is unpredictable. Note that the predictability of traffic demand is only relevant to \sys's tenant-queue binding module, which affects the performance of work conservation. Thus, we mainly focus on evaluating of work conservation when handling unpredictable demands. 

We use the same set of tenants as in \S \ref{sec:evaluation:predictable}. To generate unpredictable traffic demands, each client requests flow transmissions from randomly selected servers. Flow sizes are sampled from the empirical datacenter workloads in deployed datacenters~\cite{conga}. When the current flow finishes, a client randomly switches between being active (\ie requesting a new flow transmission) or dormant (\ie sleeping for a random period of time between 0 to 1 second before requesting a new flow transfer). 

Figure~\ref{fig:mapreduce:a} illustrates the runtime core link utilization over a one-minute measurement period. We measure the aggregated link utilization from all tenants at the granularity of 0.1 second. As illustrated in Figure~\ref{fig:mapreduce:a}, in spite of unpredictable demands, under-utilized cases are rare, rendering over $91\%$ average link utilization (plotted Figure~\ref{fig:mapreduce:b}). This is because that \sys does not rely on good TM estimation to achieve work conservation. Instead, for any D-tenant (tenant with a dedicated queue), its VMs can burst traffic with arbitrary communication patterns, allowing them to effectively grab possible spare bandwidth. As long as one VM pair from all D-tenants is high-demanded, it can drive the core link to full utilization. Mathematically, the probability that all VM pairs from D-tenants have insufficient demands is low. In particular, assuming each VM pair independently determines to be either active or dormant with equal probability during a small time interval, the probability that the core link observes insufficient demands in the small interval \footnote{Given a small interval (\eg sub-millisecond), a small flow transmission may be considered as sufficient demand.} is $(\frac{1}{2})^{\mathcal{N}}$, where $\mathcal{N}$ is the number of VM pairs from all D-tenants. Thus, demand unpredictability has minor effects on work conservation.

We further plot the average core link utilization for different lengths of control interval in Figure~\ref{fig:mapreduce:b}. For predictable demand trend, \sys achieves perfect work conservation as long as the length of control interval is comparable with how long the trend lasts. For unpredictable trend, the utilization drops slightly as the length of control interval increases. 

The takeaway of this evaluation is that to achieve good work conservation, \first \sys does not require perfect demand prediction and \second it is sufficient to perform tenant-queue allocation at coarse time granularity (\eg seconds). \emph{Thus, \sys's dynamic queue-tenant binding module does not need to react quickly enough to capture traffic bursts, which significantly reduces the stress for large scale datacenter deployment in practice. }

\parab{Fairness.} We now consider the benefits of enabling \es-like rate allocations for the tenants without  dedicated queues. First, it improves the fairness for sharing the spare bandwidth since both tenants in the shared queue and tenants with dedicated queues are able to utilize such bandwidth. Further, it slightly improves the overall link utilization by leveling up those under-utilized cases shown in Figure~\ref{fig:mapreduce:a}.

\subsection{Tenant Application Benefits}\label{sec:evaluation:FCT}
\begin{figure}[t]
	\centering
	\mbox{
		\subfigure[Enterprise workload]{\includegraphics[scale=0.38]{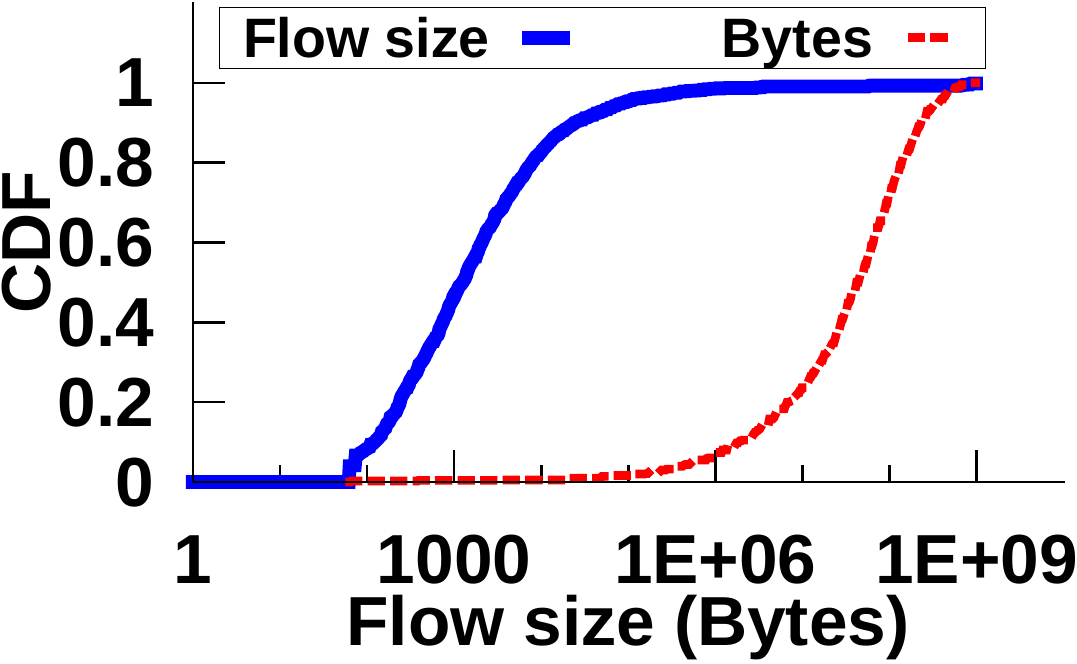}}
		\subfigure[Data-mining workload]{\includegraphics[scale=0.38]{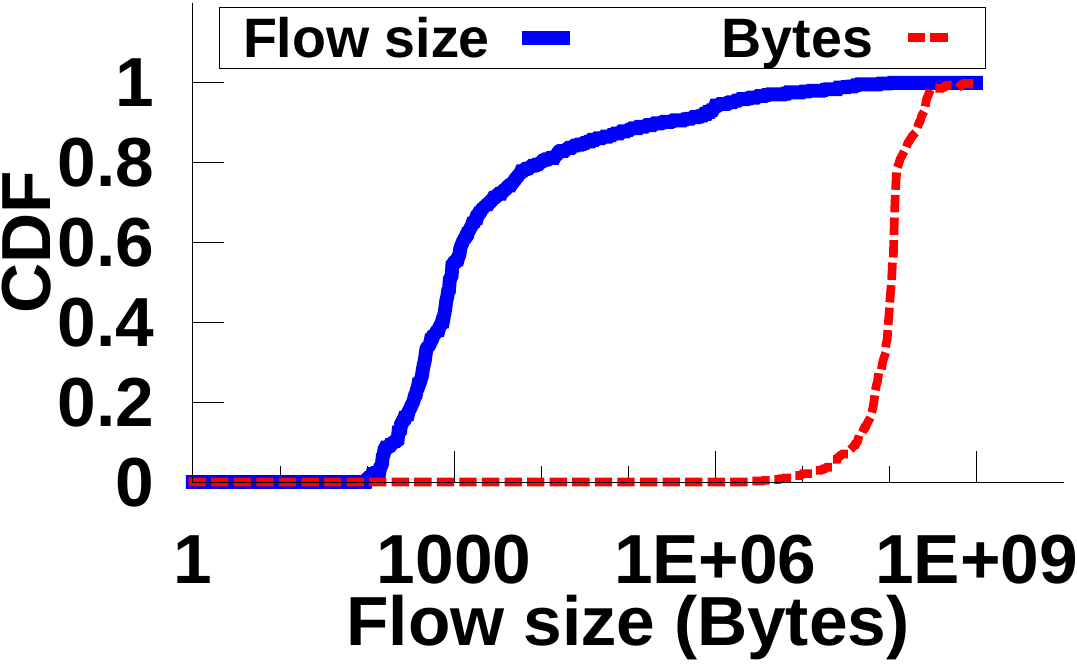}}
	}
	\caption{Empirical traffic distributions used for measuring FCTs. 
		The Bytes CDF shows the distribution of traffic bytes 
		across different flow sizes.} 
	\label{fig:flow_dis}
\end{figure}

Given the desirable property in \S \ref{sec:evaluation:WCBG}, \sys can benefit tenant applications by significantly reducing their flow completion times (FCTs). In this section, we demonstrate \sys's edges over ElasticSwitch~\cite{elasticswitch}, Trinity~\cite{trinity} as well as the static reservation for improving FCTs. Among all embedded tenants, we consider one tenant $\mathbf{T}$ with $10$ VMs evenly distributed in two racks. Tenant $\mathbf{T}$ has $94$ Mbps guaranteed bandwidth on the core link. We consider the shuffle phase of MapReduce jobs where a client requests flow transfers from all servers (recall that a VM runs both the client and server program). The flow sizes, illustrated in Figure~\ref{fig:flow_dis}, are sampled from empirically observed traffic patterns in two deployed datacenter traces~\cite{vl2} and~\cite{conga}. Each client requests a new flow once the previous one is finished, indicating that $\mathbf{T}$ has  higher demands.

In the experiment, we create different datacenter fabric loads by varying the bandwidth guarantees required by background tenants (\ie the tenants competing with $\mathbf{T}$ on the core link). The load is computed as the ratio of total guaranteed bandwidth from background tenants to the core link capacity. The results for using the enterprise datacenter workload~\cite{conga} are illustrated in Figure~\ref{fig:fct} (results for using the data-mining workload~\cite{vl2} are similar and we omit them for brevity). Because of the efficient resource utilization, \sys greatly reduces FCTs compared with both \es~\cite{elasticswitch} and the static bandwidth reservation. Such improvement is even more significant for smaller fabric loads. In spite of its improvement over static reservation, \es~\cite{elasticswitch} has a non-trivial performance degradation from \sys (up to $2\times$ long FCTs) even if it adopts very  aggressive RA to probe available bandwidth (scarifying bandwidth guarantees~\cite{elasticswitch}). We are aware that \es's performance depends on parameter settings and system tuning. Our self-implemented \es prototype uses the default parameter setting in its paper. We do not further plot the results of Trinity~\cite{trinity} since \es with aggressive RA has roughly the same performance with Trinity in terms of bandwidth utilization (\S \ref{sec:appendix}), whereas Trinity has reordering and starvation issues.

We note that \sys is orthogonal to the approaches~\cite{dctcp, d2tcp, pdq, pfabric,pias} that reduce FCTs by creating more efficient transport protocols. Rather, \sys is an approach focusing on allocating network resources for tenants, which are agnostic to both transport protocols and tenant applications. 

\subsection{\sys in Large Scale}\label{sec:evaluation:large_scale}
In this section, we evaluate \sys in large scale. In particular, we shed light on the extent of switch queue scarcity (compared with the number of tenants) in large scale datacenters. Further, we show \sys's benefit for providing tenants more bandwidth than their  guarantees and improving link utilization efficiency in large scale datacenters. We consider a three-layer multi-rooted tree topology with 1024 servers and 100 VMs per server, for a total of $100$ thousand VMs. The network interface of each server is $10$ Gbps and the switch port capacity is $40$ Gbps. The network topology is constructed based on the $k{=}16$ fattree~\cite{fattree} topology. By disabling certain links and switches, we can create a topology with different over-subscription ratios.

\subsubsection{The Extent of Switch Queue Scarcity}\label{sec:evaluation:queue_scarcity}
To be consistent with the production datacenters~\cite{seawall, oktopus}, the number of VMs requested by each tenant follows an exponential distribution with mean $49$. The bandwidth guarantee of each VM is randomly sampled from five values $10$ Mbps, $50$ Mbps, $100$ Mbps, $200$ Mbps and $300$ Mbps to better represent various bandwidth requirements from tenants. In the experiment, we keep embedding tenants until either network resources or computation resources are fully reserved, \ie the datacenter operates at $100\%$ load.  To do a stress test for queue scarcity, we assign more weight to $c_q$ in Algorithm~\ref{alg:placement}. We test three different over-subscription ratios $1:1$, $4:1$ and $16:1$.

The tenant placement results are tabulated in Table~\ref{tab:vm_placement}. Overall, the extent of queue scarcity is moderate, counterintuitive to the common assumption~\cite{faircloud}. For instance, only ${\sim}4\%$ switch ports are overloaded in the 1:1 over-subscribed topology. Among the over-utilized ports, the largest number of tenants handled by a single port is $12$, slightly higher than the total number of queues. From the tenants' perspective, two thirds of them are assigned dedicated queues throughout their lifetime due to the lack of queue contention, \ie on any link of their TRs, the number of competing tenants is less than $8$. ${\sim}90\%$ of all tenants can have  dedicated queues, either permanently or opportunistically, in any control interval, indicating that only a small fraction of tenants need to run rate allocations at hypervisors. After tenant placement, we assign tenants \textsf{dscp} values to analyze the \textsf{dscp} usage concern mentioned in \S \ref{sec:wfq_configuration}. \textsf{dscp} $0$ is reserved for tenants in shared queues. For each tenant with dedicated queues, we greedily assign it the next non-conflicting \textsf{dscp} value. It turns out that  $64$ \textsf{dscp} values are sufficient even for the fully reserved datacenter.

We further emulate the processes of tenant arrival and departure. The tenant arrival is modeled by a Poisson Process with rate $\lambda$ and the lifetime of each tenant is a constant, similar to~\cite{tag}. By varying $\lambda$, we tune the datacenter load. As the datacenter load drops, the queue scarcity is mitigated as well. When the load is less than ${\sim}60\%$, all tenants are permanently assigned dedicated queues. This demonstrates that our  tenant placement module effectively spreads tenants across available switch queues to relieve queue contention.

The takeaway for this evaluation is that in reality, the problem of queue scarcity is moderate. By performing dynamic tenant-queue binding, \sys can effectively handle such scarcity in large scale datacenters.

\subsubsection{\sys's Performance in Large Scale}
\begin{table}
  \small
  \begin{tabular}{| c  || c || c || c || c || c | }
    \hline
    O.  R. & $\mathbf{R}_{\mathbf{N}_L {<}9}$ & $\mathbf{R}_{\mathbf{N}_L {\in}[9,12]}$  & $\mathbf{R}_{\mathbf{N}_L {>} 12}$  & $\mathbf{R}_{\mathbf{N}_D}$ & $\mathbf{R}_{\mathbf{N}_I}$\\ \hline
    $1:1$ & $96.7$ & $3.26$ & $0$ & $66.7$ & $90.4$ \\ \hline
    $4:1$ & $95.1$ & $4.88$ & $0$ & $67.2$ & $90.1$ \\ \hline
    $16:1$ & $92.1$ & $7.89$ & $0$ & $66.7$ & $90.6$ \\
    \hline
  \end{tabular}
       \caption{Tenant placement results in a large scale datacenter. 
       $\mathbf{R}_{\mathbf{N}_P {<}9}$ is the percentage of ports serving less than $9$ tenants. 
       $\mathbf{R}_{\mathbf{N}_L {\in}[9,12]}$ and $\mathbf{R}_{\mathbf{N}_L {>} 12}$ 
       have similar definitions.  $\mathbf{R}_{\mathbf{N}_D}$ is the percentage of 
    tenants permanently assigned a dedicated queue and $\mathbf{R}_{\mathbf{N}_I}$ 
    is the percentage of tenants assigned a dedicated queue, 
    either permanently or opportunistically, 
    in any control interval.}\label{tab:vm_placement}
\end{table}

In this section, we evaluate \sys's performance in large scale datacenters. We show that \sys produces significant throughput gain for tenants over their guaranteed bandwidth and achieves efficient link utilization. We develop a simulator incorporating \sys's tenant placement module and dynamic queue allocation algorithm. Due to the scalability of accurately simulating detailed packet-level commutations involving billions of VM pairs, our simulator does not further study the performance of \es~\cite{elasticswitch} and Trinity~\cite{trinity} since both of them require GP which depends on accurately modeling packet-level communications. Instead, our simulator focuses on modeling tenant-level throughput, assuming tenant applications can use the available bandwidth with arbitrary communication patterns. The experiment is performed on the $16:1$ over-subscribed topology since it has the highest level of queue scarcity compared with other settings. Meanwhile we still consider the tough scenario where the datacenter operates at full load, \ie resources are fully reserved. We define the inactive ratio $r_{in}$ as the percentage of low-demanded tenants. 

\parab{Throughput Gain.} The throughput gain for a tenant is defined as the ratio of its actual achieved throughput to its guaranteed bandwidth. For simplicity, we assume the throughput gain for tenants in shared queues is $1$ (no gain). For a tenant $\mathbf{T}$ with dedicated queues, its bandwidth gain on different links of its TR may vary since the actual traffic demands on each link vary. We quantify the throughput gain of $\mathbf{T}$ as the smallest  bandwidth gain that $\mathbf{T}$ obtains on any link of its TR. Thus, our experiment shows the worst-case throughput gain for $\mathbf{T}$ when the link with the smallest bandwidth gain is the bottleneck. 

\begin{table*}[]
	\centering
	\resizebox{0.98\textwidth}{!}{
		\begin{tabular}{|c||c||c||c||c||c||c||c|}
			\hline
			& \begin{tabular}[c]{@{}c@{}}SecondNet~\cite{secondnet}, \\ Oktopus~\cite{oktopus}, TIVC~\cite{only} \end{tabular} 
			& CloudMirror~\cite{tag}                                                            
			& \begin{tabular}[c]{@{}c@{}} \es~\cite{elasticswitch}, \\ Trinity~\cite{trinity}  \end{tabular}  
			& \begin{tabular}[c]{@{}c@{}} EyeQ~\cite{eyeq}, \\ GateKeeper~\cite{gatekeeper}  \end{tabular}                                                         
			& Silo~\cite{silo}                                                              
			& QJump~\cite{qjump}                                                            
			& \sys \\ \hline
			\begin{tabular}[c]{@{}c@{}}BG\end{tabular}                
			& \textbf{Yes}     
			& \textbf{Yes}                                                            
			& Tradeoff~\cite{elasticswitch}      
			& \textbf{Yes}                                                           
			& \textbf{Yes}                                                               
			& \textbf{Yes}                                                              
			& \textbf{Yes}  \\ \hline
			\begin{tabular}[c]{@{}c@{}}WC\end{tabular}                   
			& No
			& No                                                            
			& Tradeoff~\cite{elasticswitch}
			& \textbf{Yes}                                                           
			& No                                                                
			& \textbf{Yes}                                                              
			& \textbf{Yes}  \\ \hline
			\begin{tabular}[c]{@{}c@{}}Multi-tenant \\ isolation \& placement\end{tabular} 
			& \textbf{Yes}     
			& \textbf{Yes}                                                            
			& No            
			& \textbf{Yes}                                                           
			& \textbf{Yes}                                                               
			& No                                                              
			& \textbf{Yes}  \\ \hline
			\begin{tabular}[c]{@{}c@{}}Others\end{tabular}                  
			& \textbf{None}    
			& \begin{tabular}[c]{@{}c@{}}Application  \\ driven\end{tabular} 
			& \begin{tabular}[c]{@{}c@{}}TM estimation; \\  Starvation \& reordering~\cite{trinity}\end{tabular}         
			& \begin{tabular}[c]{@{}c@{}}Non-congested \\ network core\end{tabular} 
			& \textbf{None} 
			& \textbf{None} 
			& \textbf{None} \\ \hline
	\end{tabular}}
	\caption{Property comparison with closely related works. ``BG'' and ``WC'' 
		mean bandwidth guarantee and work conservation.} \label{tab:related}
	\vspace{-0.1in}
\end{table*}
\normalsize

\begin{figure}[t]
	\centering
	\mbox{
		\subfigure[Throughput gain\label{fig:large_scale:a}]{\includegraphics[scale=0.25]{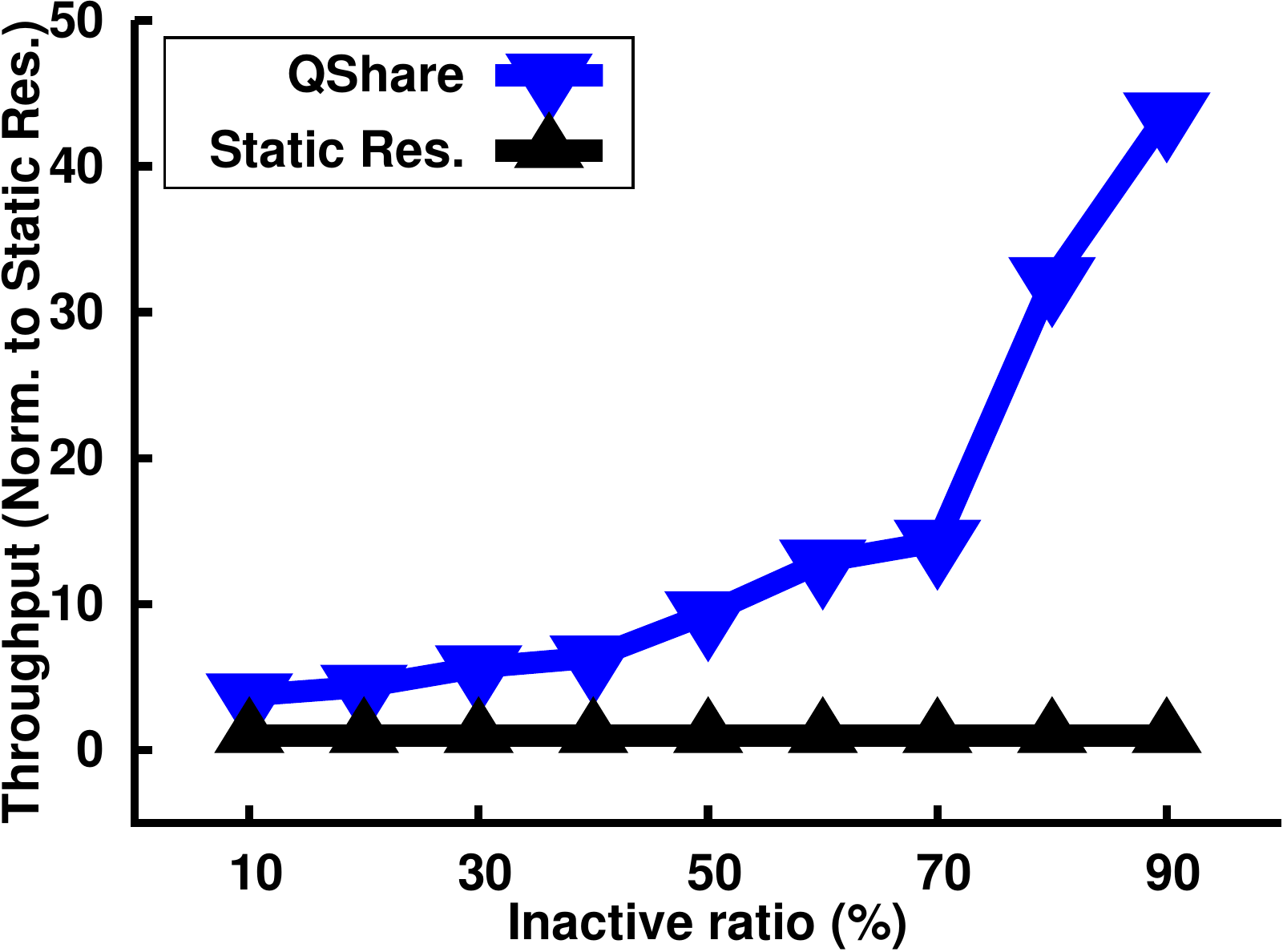}}
		\subfigure[Link utilization\label{fig:large_scale:b}]{\includegraphics[scale=0.25]{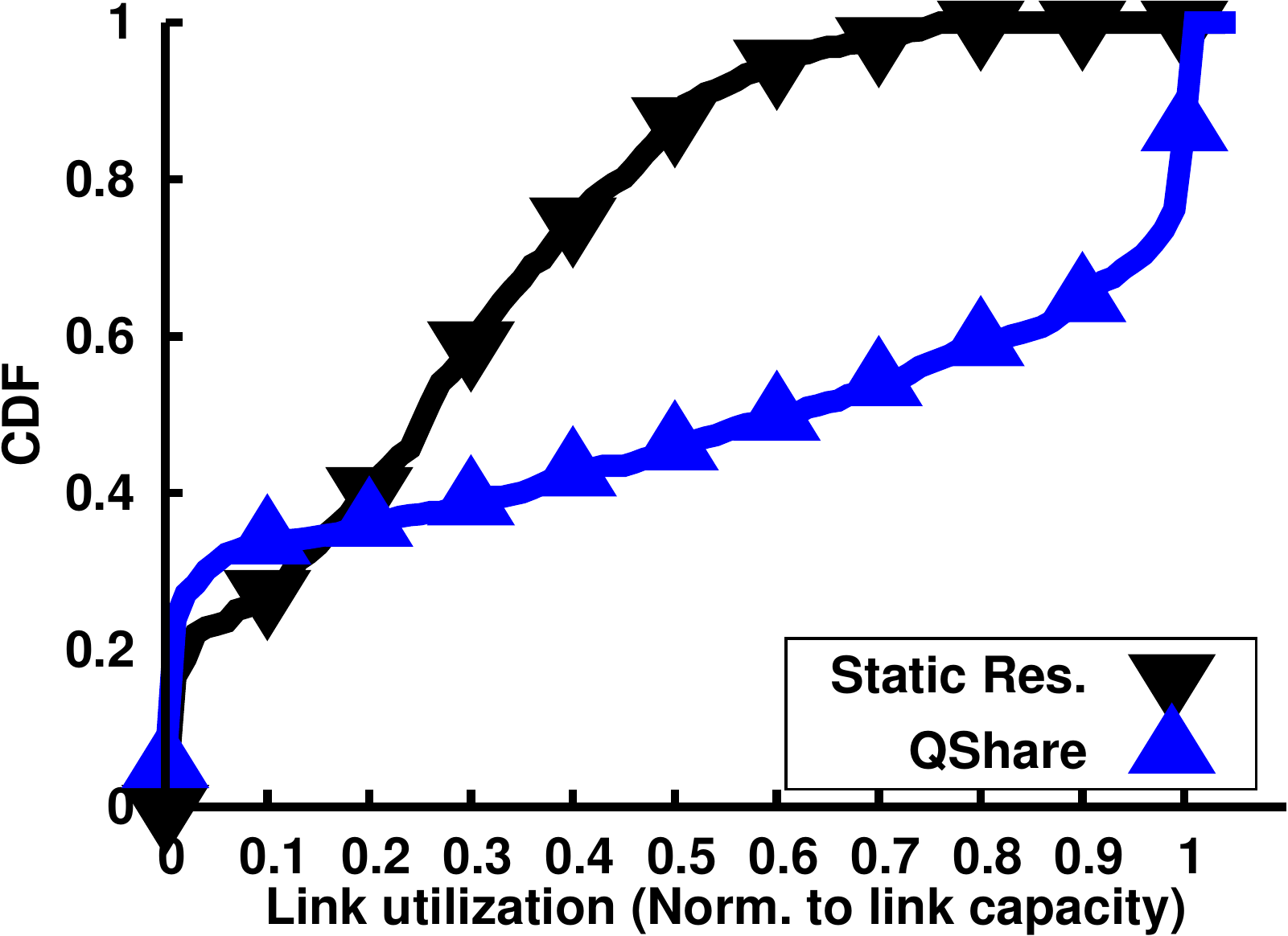}}
	}
	\caption{\sys's performance in large scale datacenters. Figure \ref{fig:large_scale:a} plots the tenants' average throughput gain over the static reservation with various inactive ratios. Figure \ref{fig:large_scale:b} shows the CDFs of normalized link utilization for \sys and static reservation.
	} 
	\label{fig:large_scale}
\end{figure}

Figure~\ref{fig:large_scale:a} illustrates the average throughput gain given varying inactive ratios. Overall, \sys produces significant throughput gains (\eg over $3\times$ for all inactive ratios) over bandwidth guarantees. The throughput gain increases dramatically (up to ${\sim}50$) as the inactive ratio increases, demonstrating that \sys can effectively utilize spare bandwidth.

\parab{Utilization Efficiency.} A natural benefit of work conservation is that \sys can improve link utilization efficiency, \ie more links are operating at high utilization, which ultimately reduces the cost of over-provisioning network bandwidth. Specifically, consider that tenant $\mathbf{T}$'s throughput gain allows it to receive an extra $100$Mbps bandwidth besides its guaranteed bandwidth. This extra bandwidth will distribute among the links of $\mathbf{T}$'s TR, driving these links to higher utilization. Without loss of generality, we consider a communication pattern spreading the throughput gain across $\mathbf{T}$'s links proportionally to $\mathbf{T}$'s guaranteed bandwidth on these links. As the throughput gain is obtained as the minimal bandwidth gain among all links on  $\mathbf{T}$'s TR, this distribution will not drive any link to over $100\%$ utilization. 

Figure~\ref{fig:large_scale:b} plots the CDFs of normalized link utilization (to the link capability) in the datacenter given $r_{in}{=}0.5$. The results show that \sys achieves better  efficiency in link utilization than static reservation. For instance, with \sys, half of the links' utilization is over ${\sim}60\%$ compared with ${\sim}25\%$ in static reservation; ${\sim}14\%$ links are fully utilized with \sys compared with $0$ percentage in static reservation. These bottleneck links show that \sys has driven the  network to the maximum possible utilization, \ie achieving work conservation. 

\subsection{System Properties}\label{sec:evaluation:benchmark}
In this section, we report the following system properties to demonstrate \sys's scalability. 
 
\parab{Switch Configuration.}
The network action container (\S \ref{sec:wfq_configuration}) executes switch configuration commands in a batch. The latency for configuring queues on all 48 ports of our legacy switch is less than $50$ms, and the configurations on different switches are parallelized using multi-threading. For OpenFlow switches, configurations can be finished almost in real time via SDN controllers such as OpenDayLight~\cite{opendaylight}. Thus, even in large scale datacenter with thousands of switches, the overall configuration latency is negligible, compared with the length of control intervals (\S \ref{sec:evaluation:unpredictable}).

\parab{CPU Overhead.} 
The major CPU overhead is contributed by the kernel module on hypervisors (\S \ref{sec:implementation}), which is affected by traffic volume. At the full NIC speed (about $940$ Mbps), we measure ${\sim}3\%$ CPU overhead on our servers (shipped with a quad-core Intel 2.8 GHz CPU).

\parab{Tenant Placement.} In the large scale network topology in \S\ref{sec:evaluation:large_scale}, the average time for figuring out the most desired TR for a 
tenant request is ${\sim}60$ms whereas the worst case takes no more than $100$ms. 

%% file: related.tex
\section{Related Work}\label{sec:related}
Table~\ref{tab:related} summarizes the properties of some closely related work. SecondNet~\cite{secondnet}, Oktopus~\cite{oktopus}, and TIVC~\cite{only} provide static, non work-conserving bandwidth guarantees. EyeQ~\cite{eyeq} and GateKeeper~\cite{gatekeeper} achieve work-conserving bandwidth guarantees only if the network core is congestion-free, which may be not true for many datacenters~\cite{high_ava, imc2010, dc_mesaure}. \es~\cite{elasticswitch} relies on challenging traffic matrix estimation and has a tradeoff between providing accurate bandwidth guarantees and being sufficiently work-conserving. Trinity~\cite{trinity} improves \es's work-conservation in static context via in-network priority queuing. However, it has starvation and packet reordering issues. Although Silo~\cite{silo} and QJump~\cite{qjump} can provide both bandwidth and in-network latency guarantee, Silo is not work-conserving and QJump lacks the tenant placement and isolation. 

Using switch queues has been proposed before. For instance, vShaper~\cite{vshaper} proposes to virtualize the physical queues to mimic the traffic shaping behavior of more queues, without considering bandwidth guarantees. pFabric~\cite{pfabric}, QJump~\cite{qjump} and PIAS~\cite{pias} instead use priority queues to achieve low latency, although pFabric requires  new hardware support, such as P4~\cite{p4}.

FairCloud~\cite{faircloud} proposes several models (or design principles) for sharing the network resources in datacenter. \sys's design follows the PS-P model, which, in theory, supports both work conservation and bandwidth guarantees simultaneously.

The bandwidth guarantees defined in the hose model can be enforced either at the level of per-tenant (\eg \cite{end-to-end, oktopus}) or at the level of VM pairs, as proposed in \cite{elasticswitch, tag,trinity}. Generally, \sys enforces per-tenant guarantees.  However, for tenants without dedicated queues, their bandwidth guarantees need to be enforced through VM-pair guarantees. 

%% file: conclusion.tex
\section{Conclusion}\label{sec:conclusion}
This paper presented \sys, the first comprehensive in-network solution enabling work-conserving bandwidth guarantees in multi-tenant datacenters. At its core, \sys's tenant placement module provides accurate bandwidth guarantees, and its tenant-queue binding module dynamically assigns high-demand tenants dedicated switch queues to achieve work conservation. We implement a prototype of \sys, and perform extensive evaluations on physical testbed and via simulations to validate \sys's design goals. The results show that \sys improves state-of-the-art solutions in two aspects: \first it does not rely on challenging traffic matrix prediction to achieve good performance and \second it eliminates the tradeoff of providing good bandwidth guarantees and being work conserving without raising starvation or packet reordering issues. Finally, \sys imposes small system overhead. 

%% file: appendix.tex
\section{APPENDIX}\label{sec:appendix}
In this section, we revisit the experiments in \S \ref{sec:in_network_support}. We would like to share our experience of  experimenting with \es~\cite{elasticswitch} and Trinity~\cite{trinity} to validate our motivation of proposing \sys. We implement a prototype of both \es and Trinity based on the designs in their papers. 

We first quantify \es's tradeoff of providing accurate bandwidth guarantees and being sufficiently work-conserving. Since the traffic matrix is unknown a priori, \es needs to allocate the bandwidth to each VM pair based on network condition probing. As a result, conservative allocation may result in bandwidth waste, especially when the total guaranteed (reserved) bandwidth is smaller than the link capability, whereas aggressive allocation may affect other tenants' guarantees, especially when large numbers of VM pairs are competing one congested link. 

\begin{figure}[h]
	\centering
	\mbox{
		\subfigure[Conservative probing\label{fig:tradeoff:a}]{\includegraphics[scale=0.25]{work_tradeoff.pdf}}~~
		\subfigure[Aggressive probing\label{fig:tradeoff:b}]{\includegraphics[scale=0.25]{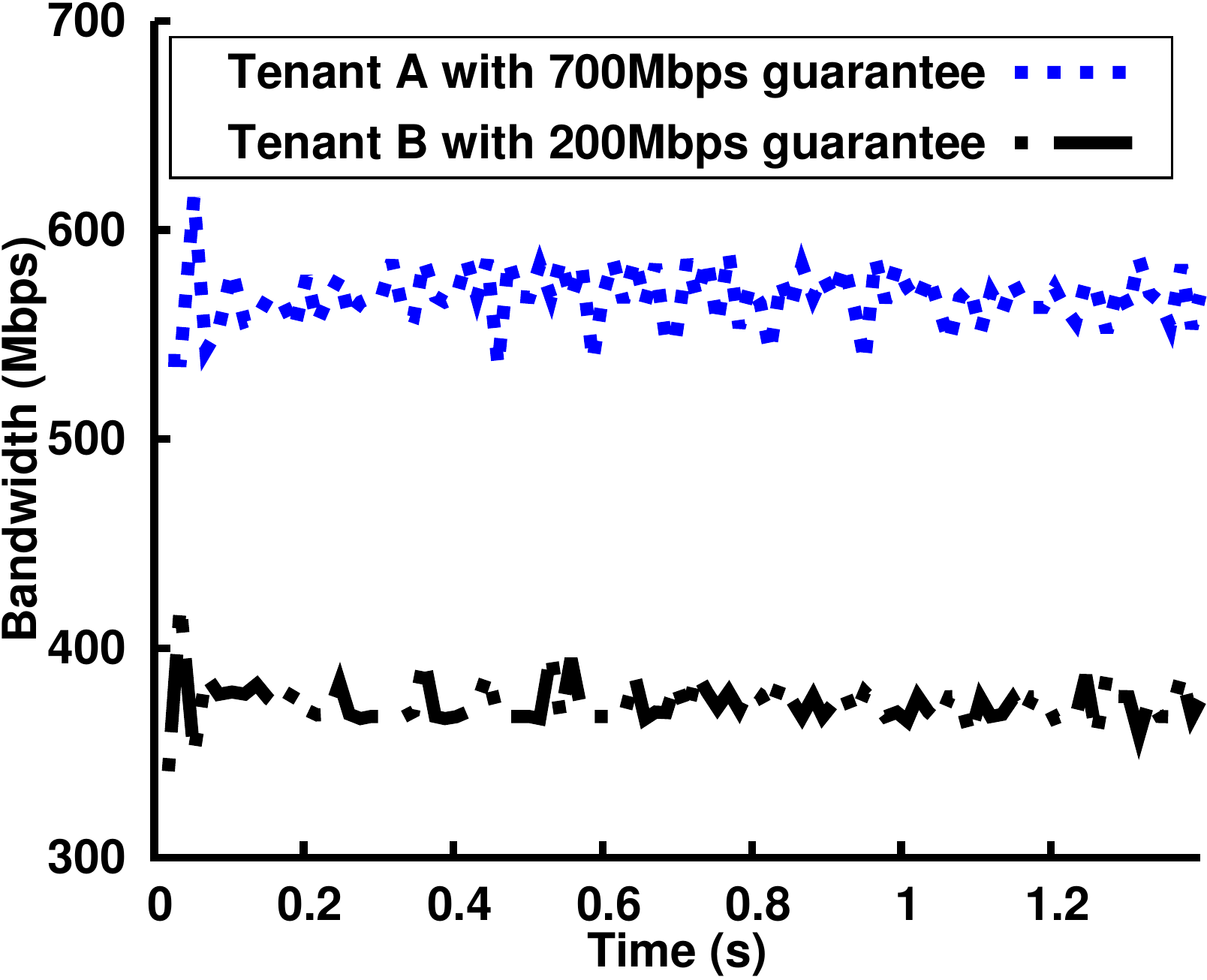}}
	}
	\caption{Quantify the tradeoff between providing accurate bandwidth guarantees and being sufficiently work-conserving for \es~\cite{elasticswitch}. 
	}
	\label{fig:tradeoff}
\end{figure}

\parab{Conservative Allocation.}
During conservative allocation, \es~\cite{elasticswitch} uses the following three mechanisms: \first Headroom: leaving a gap between the link capacity and the maximum offered guarantees on any link; \second Hold-Increase (HI): delaying the rate increase after each congestion event and \third Rate-Caution (RC): being less aggressive to increase rates once the current rates are above the guarantees.  Please refer to \cite{elasticswitch}  for the details of each mechanism. 
We use the following experiment to show the bandwidth waste caused by the conservative allocation. Consider the case where both tenant \textsf{A} and \textsf{B} adopt the same symmetric hose model, in which each VM is guaranteed $50$ Mbps bandwidth. Thus, both tenants have $250$ Mbps guarantees  on the core link, \ie the network link is half reserved. To generate traffic, each client requests flow transmissions from randomly selected servers. Flow sizes are sampled from the empirical datacenter workloads~\cite{conga}. When the current flow finishes, a client randomly switches between being active (\ie requesting
a new flow transmission) or dormant (\ie sleeping for a random period of time between zero to one second before requesting a new flow transfer). We measure the total amount of core-link bandwidth utilized by each tenant. As illustrated in Figure~\ref{fig:tradeoff:a}, we notice a significant gap (over $300$ Mbps) between the aggregated bandwidth of \textsf{A} \& \textsf{B} and the link capacity, \ie over 60\% of the unreserved bandwidth is wasted. 

\parab{Aggressive Allocation.} To achieve aggressive allocation, we disable all those three mechanisms used in conservative allocation so that the RA module immediately increases rates on each positive feedback (lack of congestion). We consider a case where tenant \textsf{A} has $700$Mbps guarantee and \textsf{B} $200$Mbps guarantee on the core link. As shown in Figure~\ref{fig:tradeoff:b}, \es fails to guarantee \textsf{A}'s bandwidth although it drives the link to higher utilization with aggressive allocation. In fact, \es~\cite{elasticswitch} has demonstrated that bandwidth guarantees will be compromised once disabling RC and HI.

We are aware that \es's performance depends on parameter choices. Thus, in practice, network operators can boost \es's performance via system tuning. Our implementation uses the parameters specified in its paper.

\parab{Analysis.} The causes of the above performance degradation are twofold: \first the challenge of guarantees partitioning (GP) without prior knowledge of communication patterns and VM-pair demands \second the challenge of relying congestion feedback to learn real-time network bandwidth. Trinity~\cite{trinity} is effective to resolve the second cause since it does not need to learn the spare network bandwidth. Instead, it serves bandwidth-guarantee traffic in a prioritized queue so that senders can aggressively send work-conservation traffic without worrying that such aggressiveness would affect other tenants' guarantees. Thus, in terms of achieving high link utilization, Trinity and \es with aggressive RA have roughly the same performance. Both our experiments and its paper show that Trinity achieves good work-conservation in static context (the traffic matrix is known and stable). However, we also notice some practical issues such as starvation and packet reordering in our experiments with Trinity.

\begin{figure}[h]
	\centering
	\mbox{
		\subfigure{\includegraphics[width=\linewidth]{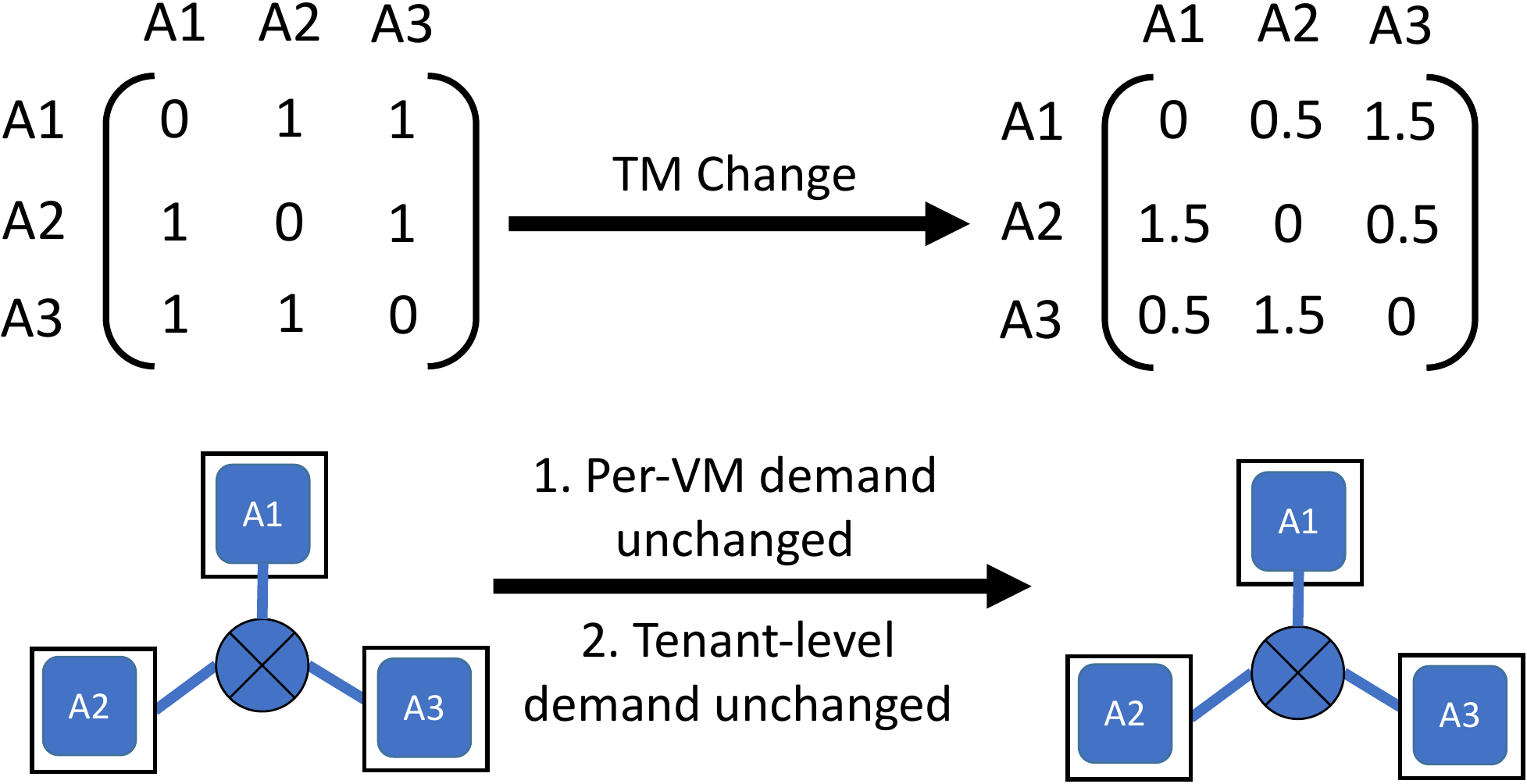}}
	}
	\caption{The GP has to re-learn each VM pair's guarantee each time the TM changes, even when both per-VM and tenant-level demand remains the same.}
	\label{fig:GP_issues}
\end{figure}

However, achieving good GP without prior knowledge still remains as an open problem. Since GP transforms each tenant's per-VM bandwidth guarantee defined in the hose model into traffic matrix (TM), it essentially complicates the model. Consider the illustrative example in Figure~\ref{fig:GP_issues}. The item $a_{ij}$ in the TM represents VM $i$'s sending demand to VM $j$. When the actual TM changes from the left pattern to the right one, GP needs to gradually learn the update via probing. However, the demand of each VM actually remains the same despite the TM change. Further, given the VM placement in Figure~\ref{fig:GP_issues} (three VMs are placed in three different hypervisors), the amount of bandwidth required for the tenant on each link also remains the same. Therefore, GP needs to do extra TM estimation even when per-VM and tenant-level demand remain the same, which essentially increases the stress of traffic demand prediction in practice.

Different from state-of-the-art solutions \es and Trinity~\cite{elasticswitch,trinity}, \sys does not rely on GP in its design. Although \sys's tenant-queue binding module does require demand prediction, it is much more lightweight than TM estimation since \sys only predicts a scalar metric for each tenant. Additionally, as shown in our evaluations (\S \ref{sec:evaluation:WCBG}), \sys does not require perfect prediction in order to achieve good work conserving bandwidth guarantees.